\documentclass[twocolumn]{aastex63}
\newcommand{\msun}{$M_{\sun}$}

\newcommand{\msunyr}{\msun\,yr$^{-1}$}

\newcommand{\mdot}{$\dot{M}$}

\newcommand{\uprime}{$u^\prime$}
\newcommand{\gprime}{$g^\prime$}
\newcommand{\rprime}{$r^\prime$}
\newcommand{\iprime}{$i^\prime$}
\newcommand{\kms}{km~s$^{-1}$}

\usepackage{graphicx}
\usepackage[caption=false]{subfig}
\usepackage{xcolor}
\usepackage{url}
\usepackage{hyperref}

\newcounter{column_number}
\shortauthors{Espaillat et al.}
\shorttitle{Radio Variability in Disks around T Tauri Stars}
\begin{document}

%\title{A Multiwavelength Variability Study of the T Tauri Star GM~Aur: Exploring the Potential for Radio Variability with Observations and Chemical Modeling}

\title{Testing the Potential for Radio Variability in Disks around T Tauri Stars with Observations and Chemical Modeling}

%% Use \author, \affil, plus the \and command to format author and affiliation 
%% information.  If done correctly the peer review system will be able to
%% automatically put the author and affiliation information from the manuscript
%% and save the corresponding author the trouble of entering it by hand.
%%
%% The \affil should be used to document primary affiliations and the
%% \altaffil should be used for secondary affiliations, titles, or email.

%% Authors with the same affiliation can be grouped in a single
%% \author and \affil call.

\correspondingauthor{Catherine C. Espaillat}
\email{cce@bu.edu}

\author{C. C. Espaillat}
\affil{Institute for Astrophysical Research, Department of Astronomy, Boston University, 725 Commonwealth Avenue, Boston, MA 02215, USA}

\author{E. Mac{\'i}as}
\affil{Joint ALMA Observatory, Avenida Alonso de C{\'o}rdova 3107, Vitacura, Santiago, Chile}

\author{J. Wendeborn}
\affil{Institute for Astrophysical Research, Department of Astronomy, Boston University, 725 Commonwealth Avenue, Boston, MA 02215, USA}

\author{R. Franco-Hern{\'a}ndez}
\affil{Instituto de Astronom{\'i}a y Meteorolog{\'i}a, Universidad de Guadalajara, Avenida Vallarta No. 2602, Col. Arcos Vallarta Sur, CP 44130, Guadalajara, Jalisco, Mexico} 

\author{N. Calvet}
\affil{Department of Astronomy, University of Michigan, 1085 South University Avenue, Ann Arbor, MI 48109, USA} 

\author{A. Rilinger}
\affil{Institute for Astrophysical Research, Department of Astronomy, Boston University, 725 Commonwealth Avenue, Boston, MA 02215, USA}

\author{L. I. Cleeves}
\affil{Astronomy Department, University of Virginia, 530 McCormick Road, Charlottesville, VA 22904, USA}

\author{P. D'Alessio}
\affil{Centro de Radioastronom{\'i}a y Astrof{\'i}sica, Universidad Nacional Aut{\'o}noma de M{\'e}xico, CP 58089, Morelia, Michoac{\'a}n, Mexico}
\affil{Deceased 2013 November 14}

\begin{abstract} 
	
A multiwavelength observing campaign of the T Tauri star (TTS) GM Aur was undertaken in 2019 December that obtained {\it Swift} X-ray and NUV fluxes, {\it HST} NUV spectra, LCOGT \uprime\gprime\rprime\iprime\ and {\it TESS} photometry, CHIRON H$\alpha$ spectra, ALMA $^{13}$CO and C$^{18}$O line fluxes, and VLA 3 cm continuum fluxes taken contemporaneously over one month. The X-ray to optical observations were presented previously. Here we present the ALMA and VLA data and make comparisons to GM Aur's accretion and X-ray properties. We report no variability in the observed millimeter CO emission. Using disk chemistry models, we show that the magnitude of the changes seen in the FUV luminosity of GM Aur could lead to variation of up to $\sim6\%$ in CO line emission and changes in the X-ray luminosity could lead to larger changes of $\sim25\%$. However, the FUV and X-ray luminosity increases must last at least 100 years in order to induce changes, which seems implausible in the TTS stage; also, these changes would be too small to be detectable by ALMA. We report no variability in the 3 cm emission observed by the VLA, showing that changes of less than a factor of $\sim3$ in the accretion rates of TTSs do not lead to detectable changes in the mass-loss rate traced by the jet at centimeter wavelengths. We conclude that typically seen changes in the FUV and X-ray luminosities of TTSs do not lead to observable changes in millimeter CO line emission or jet centimeter continuum emission.

\end{abstract}

\keywords{accretion disks, stars: circumstellar matter, 
planetary systems: protoplanetary disks, 
stars: formation, 
stars: pre-main sequence}

\section{Introduction} \label{intro}

Many pre-main-sequence stars are surrounded by circumstellar material and display the typical signatures of astrophysical accretion disks, namely mass accretion onto the central object and mass ejection via jets \citep[e.g.,][]{frank14, hartmann16}. A correlation is found between the accretion rate onto the star and the mass-loss rate of $\sim10\%$~{\mdot} \citep{cabrit90,hartigan95,ellerbroek13, watson16}.  This trend suggests a linked formation mechanism, presumably the stellar magnetic field that can both channel material onto the star as well as eject it in collimated jets along twisted field lines. In addition, while we know that young stars are highly variable across the spectrum, we do not fully understand how the variability of high-energy radiation fields results in changes in the chemistry of the disk. 

The radio wavelengths provide opportunities to look for changes in parts of the star-disk-jet system not typically studied for variability.  ALMA traces the outer disk, particularly by studying the differences in gas emission line strengths that can be traced back to the radiation field of the star.  ALMA has shown the impact of high-energy radiation fields on disk chemistry in the outer disk with an apparent link between changes in the X-ray emission and HCO$^+$ line emission \citep{cleeves17}. Centimeter-wavelength emission, such as that probed with the Karl G.\ Jansky Very Large Array (VLA), traces photoionized gas and jets \citep[e.g.,][]{macias16}. Using X-ray, UV, and centimeter data, \citet{espaillat19b} reported that the centimeter emission from the jet increases as {\mdot} increases, and that the changes in the centimeter flux are consistent with the mass-loss rate \citep{hartigan95,anglada15}.

Variability is a trademark of young ($<10$~Myr old), low-mass ($<2$~\msun), accreting pre-main sequence stars (i.e., classical T Tauri stars; CTTSs). CTTSs have magnetic fields strong enough to truncate the inner disk and lead to noncontinuous accretion of material onto the star \citep[e.g.,][]{hartmann16}.  {\mdot} is observed to be highly variable \citep[e.g.,][]{venuti14,robinson19}, down to minutes in some cases \citep{stauffer16, siwak18}. X-ray emission from the stellar corona and accretion-related processes is also known to be variable \citep{preibisch05, argiroffi11,flaccomio12,principe14,guarcello17}.  It is unclear if the variability of the high energy fields generated by accretion and the star has an impact on the surrounding disk. There have been very few multi-epoch studies of CTTSs linking variability across different wavelengths (i.e., the X-ray, UV, optical, and IR wavelengths), and those that exist have focused on regions very close to the star \citep[$<1$~au;][]{dupree12,stauffer14,flaherty14,espaillat19a}.  

In addition to accretion, CTTSs also show signs of mass ejection in the form of jets and/or winds \citep[e.g.,][]{frank14,anglada18}. The driving mechanism of these ejections of material is still not well understood, but the ejections are thought to be centrifugally launched at very close distances from the star ($<1$~au) and then collimated through magnetic fields \citep{cabrit07}. Recent radio observations have shown the presence of jets in the later protoplanetary disk stage, which suggests that jets could still affect the planet-formation process \citep{rodriguez14,macias16}.  However, radio observations of disks in their later stages can also trace photoionized gas; chromospheric activity can produce high-energy radiation that can heat and photoionize the disk, resulting in a photoevaporating wind \citep{alexander14}.  

Here we present the radio component of a comprehensive, coordinated multi-epoch, observing campaign in the X-ray to radio wavelengths of the CTTS GM~Aur. GM~Aur is a well-studied robust accretor \citep{robinson19} that is surrounded by a transitional disk \citep[i.e., an object with a large disk hole;][]{macias18,espaillat14} that also hosts a jet \citep{macias16}. \citet{espaillat21} presented the X-ray to optical part of the dataset and found evidence of rotationally modulated accretion and structure in the accretion hot spot on the stellar surface.  Here we search for variability in the $J=2-1$ CO gas lines, which could be due to changes in the gas temperature, CO photodissociation, or other chemical effects. We also test for variability in the centimeter continuum emission, which traces the jet. In Section 2, we describe the overall campaign and focus on the radio data, and in Section 3, we compare our data to disk chemistry models and search for correlations with other data in the campaign. In Section 4, we discuss our results in light of chemical modeling and previous work.

%%%%%%%%%%%%%%%%%%%%%%%%%%
\begin{figure}    
\epsscale{1.25}
\plotone{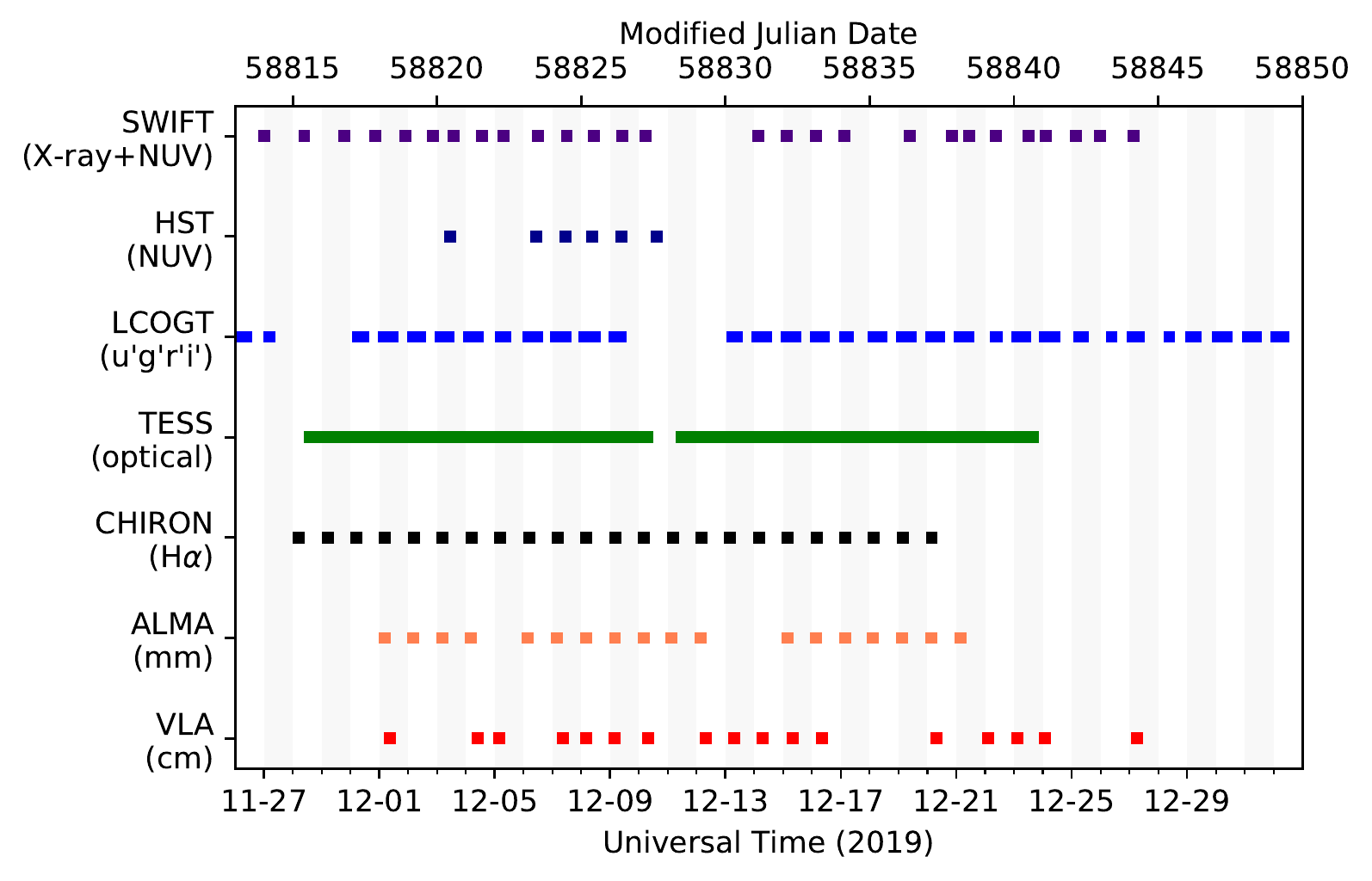} 
\caption{
Overview of our 2019 Dec campaign of GM~Aur that included {\it Swift}, {\it HST}, LCOGT, {\it TESS}, ALMA, and VLA data. When possible, data were taken once a day except for {\it TESS} (scheduled for every $\sim2$~minutes) and LCOGT (taken 5--10 times per night). Gaps are due to poor weather and/or scheduling constraints. In this work, we present the ALMA and VLA data. The other datasets were presented previously by \citet{espaillat21}.    
}
\label{fig:cadence}
\end{figure} 
%%%%%%%%%%%%%%%%%%%%%%%%%%

\section{Observations \& Data Reduction} \label{redux}

Our multiwavelength observing campaign of GM~Aur was undertaken mostly during 2019 December (Figure~\ref{fig:cadence}).  X-ray to optical data with {\it Swift}, LCOGT, {\it TESS}, {\it HST}, and CHIRON were presented in \citet{espaillat21}. Here we present data spanning millimeter to centimeter wavelengths from ALMA and VLA and provide details on the data reduction.

%TABLE
\begin{deluxetable*}{ccccc}
\tablecaption{ALMA Observations \label{Tab:ALMAobs}}
\tablehead{
\colhead{Start Time} \vspace{-0.2cm} & \colhead{End Time} & \colhead{N$_{ant}$} & \colhead{Min.--Max. baseline} & \colhead{Mean PWV} \\
\colhead{(UT)} & \colhead{(UT)} & \colhead{} & \colhead{(m)} & \colhead{(mm)}  }
\startdata
2019-12-01T04:28:45 &	2019-12-01T04:39:45 &	41 & 15.1--312.7	& 1.5	\\
2019-12-02T04:20:33 &	2019-12-02T04:31:34	&	42 & 15.1--312.7    & 1.4	\\
2019-12-03T04:28:51 &	2019-12-03T04:39:52 &	42 & 15.1--312.7	& 0.6	\\
2019-12-04T04:09:44 &	2019-12-04T04:20:45 &	41 & 15.1--312.7	& 1.0	\\
2019-12-06T03:19:35 &	2019-12-06T03:30:37 &	44 & 15.0--312.7	& 0.9	\\
2019-12-07T03:30:36 &	2019-12-07T03:41:38 &	43 & 15.1--312.7	& 0.9	\\
2019-12-08T03:52:08 &	2019-12-08T04:03:10 &	43 & 15.1--312.7	& 1.0	\\
2019-12-09T04:02:45 &	2019-12-09T04:13:47 &	44 & 15.1--312.7	& 1.1	\\
2019-12-10T03:52:29 &	2019-12-10T04:03:31 &	44 & 15.1--312.7	& 0.6	\\
2019-12-11T02:49:09 &	2019-12-11T03:00:11 &	45 & 15.0--312.7	& 0.6	\\
2019-12-12T03:23:44 &	2019-12-12T03:34:46 &	41 & 15.0--312.7	& 0.3	\\
2019-12-15T03:26:52 &	2019-12-15T03:37:55 &	43 & 15.0--312.7	& 1.1	\\
2019-12-16T03:19:30 &	2019-12-16T03:30:31 &	45 & 15.0--312.7	& 1.1	\\
2019-12-17T03:38:58 &	2019-12-17T03:50:00 &	44 & 15.0--312.7	& 0.8	\\
2019-12-18T02:42:04 &	2019-12-18T02:53:07 &	45 & 15.1--312.7	& 1.0	\\
2019-12-19T03:12:20 &	2019-12-19T03:23:22 &	42 & 15.1--312.7	& 0.8	\\
2019-12-20T03:17:49 &	2019-12-20T03:28:51 &	43 & 15.1--312.7	& 1.1	\\
2019-12-21T03:17:20 &	2019-12-21T03:28:22 &	42 & 15.1--312.7	& 1.1		 
%----------------------------------------------------------------------
\enddata
\end{deluxetable*}

%\clearpage
%TABLE 
%\movetabledown=155mm
%\begin{rotatetable}
\begin{deluxetable*}{cc|cccc|cccc}
\tablecaption{ALMA $^{13}$CO and C$^{18}$O  Line Fluxes \label{Tab:ALMAflux}}
\tablehead{
\colhead{Date}  \vspace{-0.2cm}& \colhead{rms} & \colhead{rms} & \colhead{$F_{\rm 13 CO}$}  & \colhead{$F_{\rm 13 CO}/$}  & \colhead{Ph.Cal.} & \colhead{rms} & \colhead{$F_{\rm C18O}$} & \colhead{$F_{\rm C18O}/$}  & \colhead{Ph.Cal.}\\
\colhead{} \vspace{-0.2cm} & \colhead{cont.} &\colhead{$^{13}$CO} &   \colhead{}  & \colhead{$F_{\rm cont.}$}  & \colhead{flux} 
& \colhead{C$^{18}$O} &   \colhead{} & \colhead{$F_{\rm cont.}$}  & \colhead{flux} \\
\colhead{(2019} \vspace{-0.2cm}  &  \colhead{(mJy/} & \colhead{(mJy$/$} & \colhead{(mJy} & \colhead{(10$^{-2}$)} & \colhead{(Jy)}
&  \colhead{(mJy/} & \colhead{(mJy} & \colhead{(10$^{-2}$)} & \colhead{(Jy)} \\
\colhead{UT)}  \vspace{-0.2cm} & \colhead{beam)} & \colhead{beam} & \colhead{km~s$^{-1}$)}  & \colhead{} & \colhead{}
& \colhead{beam} & \colhead{km~s$^{-1}$)}  & \colhead{} & \colhead{} \\
\colhead{}  & \colhead{} & \colhead{km~s$^{-1}$)} & \colhead{} & \colhead{} 
& \colhead{} & \colhead{km~s$^{-1}$)} & \colhead{} & \colhead{} & \colhead{} 
}
\startdata
12-01 & 0.167 &0.013	&	$4.674\pm0.038$ &	$2.685\pm0.025$	  & 0.339	 & 0.015	&	$0.986\pm0.042$ &	$0.567\pm0.024$	  & 0.348	 \\
12-02 & 0.135 &0.020	&	$5.017\pm0.056$ &	$2.736\pm0.032$	  & 0.354	 & 0.016	&	$1.094\pm0.044$ &	$0.596\pm0.024$	  & 0.365	 \\
12-03 & 0.137 &0.014	&	$4.952\pm0.042$ &   $2.756\pm0.026$	  & 0.362	 & 0.009	&	$0.975\pm0.026$ &   $0.543\pm0.015$	  & 0.364	 \\
12-04 & 0.152 &0.018	&	$4.983\pm0.049$ &	$2.732\pm0.029$	  & 0.364	 & 0.013	&	$1.053\pm0.038$ &	$0.577\pm0.021$	  & 0.368	 \\
12-06 & 0.174 &0.016	&	$5.011\pm0.049$ &	$2.753\pm0.029$	  & 0.351	 & 0.013	&	$1.071\pm0.039$ &	$0.588\pm0.022$	  & 0.359	 \\
12-07 & 0.162 &0.017	&	$4.996\pm0.053$ &	$2.713\pm0.031$	  & 0.356	 & 0.014	&	$1.194\pm0.044$ &	$0.648\pm0.024$	  & 0.360	 \\
12-08 & 0.140 &0.017	&	$5.196\pm0.040$ &	$2.777\pm0.024$	  & 0.352	 & 0.013	&	$1.163\pm0.041$ &	$0.621\pm0.022$	  & 0.362	 \\
12-09 & 0.127 &0.013	&	$5.121\pm0.037$ &	$2.761\pm0.023$	  & 0.355	 & 0.010	&	$1.106\pm0.034$ &	$0.596\pm0.019$	  & 0.351	 \\
12-10 & 0.162 &0.012	&	$5.116\pm0.047$ &	$2.745\pm0.027$	  & 0.348	 & 0.012	&	$1.062\pm0.035$ &	$0.569\pm0.019$	  & 0.354	 \\
12-11 & 0.215 &0.016	&	$4.941\pm0.042$ &	$2.741\pm0.026$	  & 0.351	 & 0.012	&	$1.107\pm0.039$ &	$0.614\pm0.022$	  & 0.344	 \\
12-12 & 0.141 &0.014	&	$4.997\pm0.047$ &	$2.803\pm0.029$	  & 0.339	 & 0.013	&	$1.117\pm0.042$ &	$0.626\pm0.024$	  & 0.337	 \\
12-15 & 0.130 &0.015	&	$4.900\pm0.036$ &   $2.799\pm0.024$	  & 0.334	 & 0.011	&	$1.110\pm0.366$ &   $0.634\pm0.021$	  & 0.344	 \\
12-16 & 0.138 &0.017	&	$4.725\pm0.049$ &	$2.692\pm0.030$	  & 0.360	 & 0.011	&	$1.117\pm0.037$ &	$0.636\pm0.021$	  & 0.350	 \\
12-17 & 0.158 &0.014	&	$4.914\pm0.046$ &	$2.767\pm0.028$	  & 0.362	 & 0.012	&	$1.027\pm0.037$ &	$0.578\pm0.021$	  & 0.353	 \\
12-18 & 0.169 &0.011	&	$4.585\pm0.040$ &	$2.682\pm0.026$	  & 0.339	 & 0.013	&	$0.941\pm0.040$ &	$0.550\pm0.023$	  & 0.346	 \\
12-19 & 0.157 &0.016	&	$4.885\pm0.047$ &	$2.711\pm0.028$	  & 0.348	 & 0.012	&	$1.023\pm0.038$ &	$0.567\pm0.021$	  & 0.357	 \\
12-20 & 0.149 &0.016	&	$5.072\pm0.049$ &	$2.787\pm0.029$	  & 0.349	 & 0.012	&	$1.118\pm0.040$ &	$0.614\pm0.022$	  & 0.369	 \\
12-21 & 0.135 &0.013	&	$5.011\pm0.041$ &	$2.773\pm0.025$	  & 0.341	 & 0.010	&	$1.063\pm0.034$ &	$0.588\pm0.019$	  &~0.365	
%----------------------------------------------------------------------
\enddata
\end{deluxetable*}
%\end{rotatetable}
%\clearpage
 
 %FIGURE 
%%%%%%%%%%%%%%%%%%%%%%%%%%
\begin{figure*}    
\epsscale{0.8}
\plotone{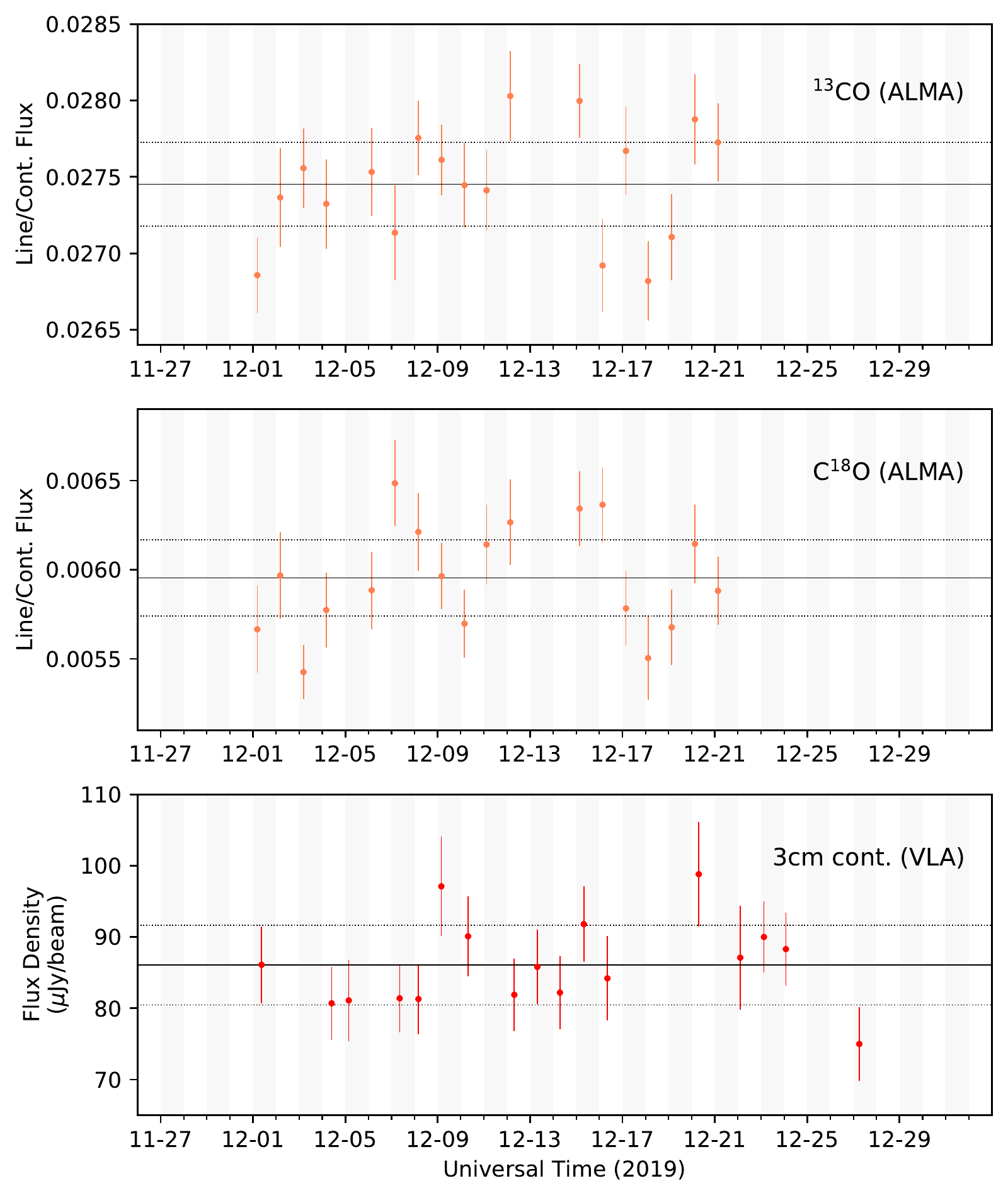}  
\caption{
Data from ALMA and VLA. We show the $^{13}$CO and C$^{18}$O line-to-continuum flux ratios and 3~cm continuum flux density. We also show the mean values (solid line) and mean uncertainties ($1\sigma$; broken lines) for the $^{13}$CO and C$^{18}$O line-to-continuum flux ratios (0.0274$\pm$0.0003 and 0.0059$\pm$0.0002, respectively) and the 3~cm continuum flux density (86.1$\pm$5.6 $\mu$Jy/beam). 
}
\label{fig:obs}
\end{figure*} 
%%%%%%%%%%%%%%%%%%%%%%%%%%

\subsection{ALMA}

ALMA observations of GM~Aur (PI: Espaillat; 2019.1.01437.S) were obtained daily between 2019 December 1 to 2019 December 21 except for three days with bad weather or scheduling constraints (Table~\ref{Tab:ALMAobs}). The same observational setup was used for all 18 visits, with two spectral windows spanning the $^{12}$CO $J=2-1$, $^{13}$CO $J=2-1$, and C$^{18}$O $J=2-1$ lines, and two dedicated to detect the continuum centered at 219.9 GHz and 231.0 GHz. The latter spectral windows overlapped the line frequencies in order to use them to self-calibrate the line emission. The on-source time for each epoch was $\sim10.1$~minutes. The quasar J0423-0120 was used as the flux and bandpass calibrator and J0438+3004 was the phase calibrator. We show the channel maps of the three transitions (combining all the epochs) in the Appendix.
The antenna configuration was not significantly modified during the observing campaign, so the {\it u,v} coverage of the visits was fairly consistent. The angular resolution of the observations is $\sim1''$, while the maximum recoverable scale is $\sim10\rlap.''7$. While these are adequate to study the gas in the disk of GM Aur \citep[R=650 au;][]{schwarz21}, the $^{12}$CO emission of GM Aur shows complex extended emission up to $12''$ from the central star  \citep{huang21}. This extended emission is not well recovered by our observations due to the lack of shorter spacings in the {\it u,v} coverage. As a consequence, small changes in {\it u,v} coverage between different epochs introduced imaging artifacts that could result in artificial variability in the measured $^{12}$CO flux. Going forward, we thus focus on the $^{13}$CO and C$^{18}$O emission lines. These lines are optically thinner than $^{12}$CO and should be coming from the disk, with little contribution from the surrounding extended material. 

The calibration of the data was performed using the automatic pipeline within \texttt{CASA} (Common Astronomy Software Applications; v. 5.6.1-8). Phase and amplitude self-calibration were performed on all visits, improving the sensitivity for the $^{13}$CO line by $\sim13\%$ and for the C$^{18}$O line by $\sim11\%$. Continuum and line images were obtained using the task \texttt{tclean}. The \texttt{mtmfs} algorithm \citep{rau11} was used to obtain 1.3~mm continuum images, assuming point-source components and a straight spectrum (scales =~0; $nterms=2$). We chose a Briggs weighting (robust =~0.5) for the continuum CLEANing. The image cubes were computed using the \texttt{hogbom} CLEANing algorithm and a Briggs weighting (robust =~0.5). No channel averaging was performed, yielding channel widths of 30.5~kHz ($\sim0.04$~\kms) for each line. 

After performing continuum subtraction using the task \texttt{uvcontsub}, a Keplerian masking script\footnote{https://github.com/richteague/keplerian$\_$mask} was used to reduce contamination from extended emission and increase the signal-to-noise ratio of the spectra and velocity-integrated maps. Final image cubes were obtained via the product of the CLEANed image cube and its corresponding Keplerian mask using the task \texttt{immath}. Velocity-integrated maps (i.e., 0th-order moment maps) were calculated via the task \texttt{immoments}. Spectra were obtained by integrating the image cubes over regions depending on the velocity of the channel set by the Keplerian masking scripts and using the flux contained within the unmasked region for each channel. Line fluxes were obtained by integrating the emission in the same unmasked region used to obtain the spectra. The uncertainties of the line fluxes were computed as $\delta_{line} = \sigma_{chan}  \Delta v \sqrt{\Omega N_{chan}}$; where $\sigma_{chan}$ is the rms of the image cube measured in the line-free channels (see Table~\ref{Tab:ALMAflux}), $\Delta v$ is the channel width of the image cube, $\Omega$ is the solid angle of the Keplerian mask in number of beams, and $N_{chan}$ is the number of channels used to create the velocity integrated maps as estimated from the applied Keplerian mask. To measure the continuum fluxes, we fit a Gaussian to a $4^{\prime\prime}\times4^{\prime\prime}$ region centered on the peak of the continuum emission. The uncertainties in the continuum fluxes were estimated from the Gaussian fit using the rms from each image (see Table~\ref{Tab:ALMAflux}). 

In order to verify the robustness of any potential variability, we first checked the flux calibration of our data. According to the ALMA Calibrator Catalogue, our flux calibrator (J0423-0120) was observed three times in Band 3 (7, 18, 19 December) and three times in Band 7 (9, 18, 19 December) during our observations and displayed roughly 10\% and 5\% variability in each band, respectively. To account for this variability in the flux calibrator, we first interpolated the Band 3 and 7 fluxes for J0423-0120 for the dates of our 18 GM~Aur observations, additionally using the nearest observations prior to and following our observations in each band. We then estimated the flux of J0423-0120 in the relevant spectral windows by assuming a constant spectral index between Band 3 and 7. Finally, we used the tasks \texttt{gencal} and \texttt{applycal} to apply these corrected fluxes to our measurement sets. 

Line fluxes are provided in Table~\ref{Tab:ALMAflux} and do not include the absolute flux calibration uncertainties. We note that the $^{12}$CO flux that we obtain from combining all the epochs is $20.86\pm0.01$~mJy \kms. In order to cancel the effects of absolute flux calibration error \citep{francis20}, we hereafter analyze the line-to-continuum ratio of the CO lines, using the continuum flux from the continuum spectral windows overlapping the different lines (see Figure~\ref{fig:obs}). We found no correlations between our measured fluxes (continuum, line, line/continuum) and different potential systematics such as the number of antennas, average T$_{sys}$ of the observation, precipitable water vapor (PWV), or beam sizes. We also tested whether variable {\it u,v} coverage could explain the small differences in percent of the flux by simulating a synthetic model image cube of the $^{13}$CO emission of the disk around GM~Aur \citep[obtained using RADMC-3D;][]{dullemond12} using the different {\it u,v} coverages of the observations. We found that differences in {\it u,v} coverage accounted for differences of only hundredths of a percent in the continuum and 0.5\% for the $^{13}$CO line, whereas the observed $^{13}$CO fluxes deviate by about 5\%.

%TABLE 
\begin{deluxetable*}{ccccccc}[ht!]  
\tablecaption{VLA Observations \& Fluxes \label{tab:vla}}
\tablehead{
\colhead{Start Time} & \colhead{End Time} & \colhead{$F_{\rm 3 cm}$} & \colhead{rms$_{\rm 3C138}$} & \colhead{rms$_{\rm 3C147}$} & \colhead{Ph. Cal.$_{\rm 3C138}$} & \colhead{Ph. Cal.$_{\rm 3C147}$}\\
\colhead{(UT)} & \colhead{(UT)} & \colhead{($\mu$Jy)} & \colhead{($\mu$Jy beam$^{-1}$)} & \colhead{($\mu$Jy beam$^{-1}$)} & \colhead{$F_{\rm 3 cm}$ (Jy)} & \colhead{$F_{\rm 3 cm}$ (Jy)}
}
\startdata
%----------------------------------------------------------------------
2019-12-01T08:56:27.0  & 2019-12-01T09:34:27.0 & $86.1\pm5.4$ & 5.00 & 5.00 & $1.2365\pm0.0004$ & $1.2422\pm0.0003$ \\
2019-12-04T09:45:12.0  & 2019-12-04T10:23:12.0 & $80.7\pm5.1$ & 5.00 & 5.00 & $1.2198\pm0.0028$ & $1.2134\pm0.0008$ \\    
2019-12-05T03:34:36.0  & 2019-12-05T04:17:39.0 & $81.1\pm5.7$ & 4.50 & 4.50 & $1.2530\pm0.0007$ & $1.2496\pm0.0003$ \\       
2019-12-07T08:28:21.0  & 2019-12-07T09:06:21.0 & $81.4\pm4.8$ & 4.50 & 4.50 & $1.2291\pm0.0011$ & $1.2302\pm0.0004$\\    
2019-12-08T03:58:42.0  & 2019-12-08T04:41:48.0 & $81.3\pm4.9$ & 4.50 & 4.50 & $1.2518\pm0.0004$ & $1.2600\pm0.0002$ \\    
2019-12-09T03:45:48.0  & 2019-12-09T04:28:51.0 & $97.1\pm7.0$ & 6.00 & 6.00 & $1.2697\pm0.0008$ & $1.2908\pm0.0008$ \\    
2019-12-10T07:34:48.0  & 2019-12-10T08:12:51.0 & $90.1\pm5.6$ & 5.00 & 4.50 & $1.2272\pm0.0003$ & $1.2304\pm0.0001$ \\    
2019-12-12T07:36:18.0  & 2019-12-12T08:14:18.0 & $81.9\pm5.1$ & 4.50 & 4.50 & $1.2253\pm0.0003$ & $1.2281\pm0.0003$ \\  
2019-12-13T07:30:39.0  & 2019-12-13T07:59:12.0 & $85.8\pm5.2$ & 5.00 & 4.50 & $1.2273\pm0.0004$ & $1.2251\pm0.0003$ \\    
2019-12-14T07:03:54.0  & 2019-12-14T07:41:54.0 & $82.2\pm5.1$ & 4.50 & 4.50 & $1.2226\pm0.0003$ & $1.2279\pm0.0002$ \\    
2019-12-15T07:49:27.0  & 2019-12-15T08:27:27.0 & $91.8\pm5.3$ & 5.00 & 4.50 & $1.2230\pm0.0008$ & $1.2300\pm0.0008$ \\    
2019-12-16T08:20:21.0  & 2019-12-16T08:58:21.0 & $84.2\pm5.9$ & 5.00 & 4.50 & $1.2304\pm0.0023$ & $1.2549\pm0.0005$ \\
2019-12-20T07:18:36.0  & 2019-12-20T07:56:39.0 & $98.8\pm7.3$ & 4.50 & ... 	& $1.2388\pm0.0009$ & ... \\   
2019-12-22T02:27:09.0  & 2019-12-22T03:10:12.0 & $87.1\pm7.3$ & ...  & 4.50 & ... 				& $1.2556\pm0.0001$ \\   
2019-12-23T03:02:03.0  & 2019-12-23T03:45:09.0 & $90.0\pm5.0$ & 4.50 & 4.00 & $1.2572\pm0.0002$ & $1.2546\pm0.0001$ \\   
2019-12-24T01:51:06.0  & 2019-12-24T02:34:09.0 & $88.3\pm5.1$ & 4.50 & 4.50 & $1.2699\pm0.0003$ & $1.2681\pm0.0002$ \\    
2019-12-27T06:09:24.0  & 2019-12-27T06:47:24.0 & $75.0\pm5.2$ & 4.50 & 4.50 & $1.2350\pm0.0002$ & $1.2401\pm0.0002$    
%----------------------------------------------------------------------
\enddata
\end{deluxetable*} 

\subsection{VLA}

VLA continuum observations of GM~Aur at 3~cm (X~band) were taken as part of Project 19A-492 (PI: Espaillat; Table~\ref{tab:vla}). Each of the 17 observations had $\sim40$~minutes of on-source time. The VLA was in its B configuration, and all antennas remained in the same position for each observation. The observations were taken within the same 2~hour LST window, ensuring that the {\it u,v} coverage was consistent. Two different quasars (3C138 and 3C147) were used as the flux and bandpass calibrators in order to ensure an accurate flux calibration. J0414+3418 was used as the phase calibrator.

These data were calibrated using the VLA pipeline in CASA (version 5.1.0). The data were inspected and data with irregular phases or amplitudes were flagged. CLEANed images were then obtained using \texttt{tclean} in CASA (version 5.3.0). We used the \texttt{mtmfs} algorithm with $nterms=2$ and pointlike components (scales =~0). Integrated continuum flux densities were measured by fitting a Gaussian to the convolved images. We obtained two images of GM~Aur for each date of observing: one using each of the two flux calibrators. We fit a Gaussian to each of the images to measure the flux and uncertainty, then averaged the two fluxes for each date using the uncertainty as the weighting. We report the weighted average 3~cm flux density (Table \ref{tab:vla}) and use these values in Figure~\ref{fig:obs}.

We note that there were technical issues with the data for 3C138 on 20 December and 3C147 on 22 December, so on those days, the reported flux density is based on only one available calibration source. We note that the uncertainties listed in Table \ref{tab:vla} do not include the expected 5\% absolute flux uncertainty of the VLA at 3~cm, but these flux calibration uncertainties are included when plotting the data in Figure~\ref{fig:obs}.

\section{Analysis \& Results} 

\subsection{Observational Results}

Here we study the $^{13}$CO and C$^{18}$O emission lines observed by ALMA and the 3~cm continuum flux densities measured by the VLA.  We explore if there is any variability in these observations.

\subsubsection{$^{13}$CO and C$^{18}$O Line Emission with ALMA}
\begin{figure}    
\epsscale{1.0}
\plotone{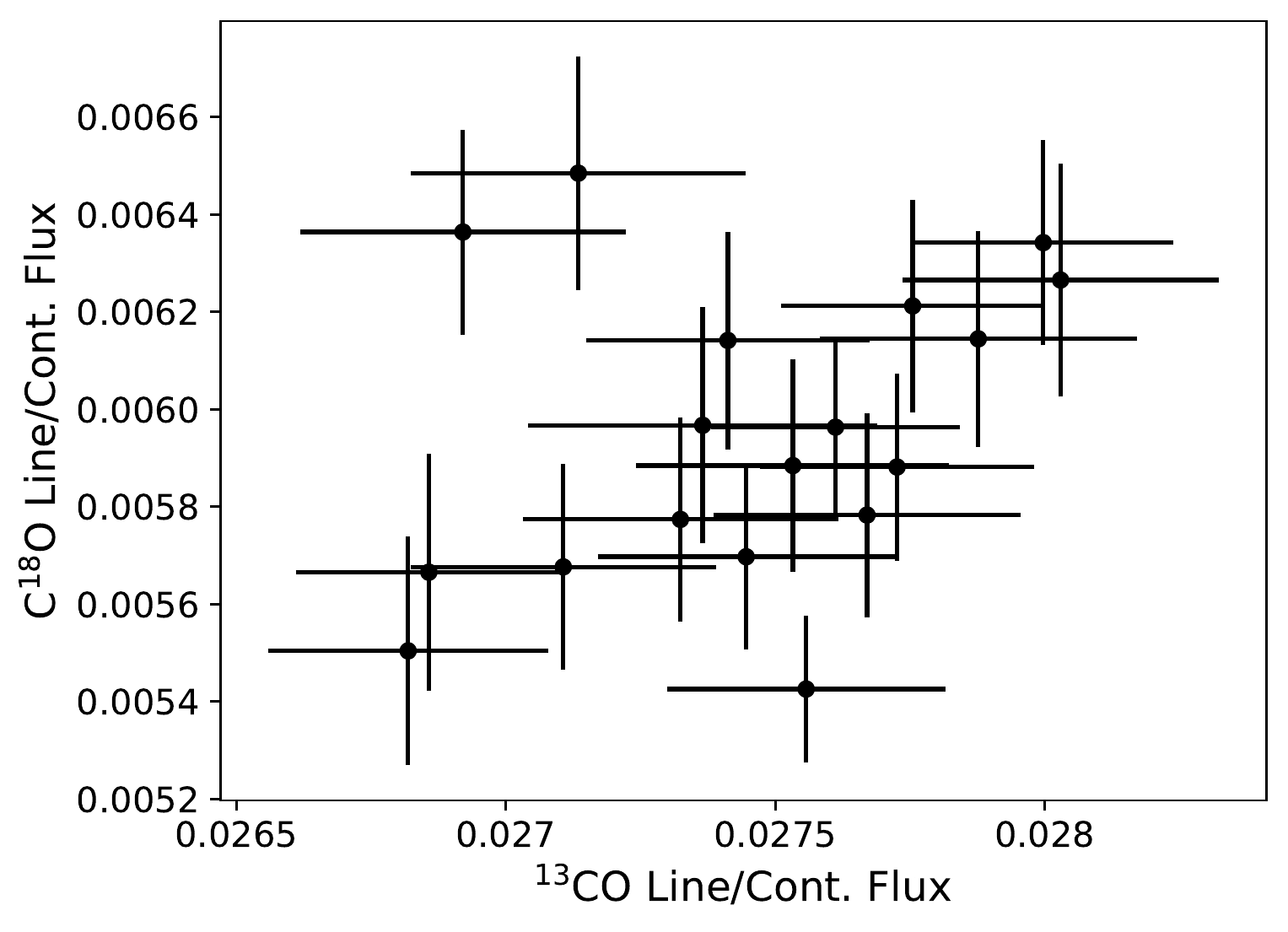}
\caption{
Comparison of the $^{13}$CO line-to-continuum ratio to the C$^{18}$O line-to-continuum ratio. There is no strong correlation ($\rho_p=0.3$, $p_p=0.2$; $\rho_s=0.4$, $p_s=0.1$; $\tau_k=0.3$, $p_k=0.1$).
}
\label{fig:corrALMA}
\end{figure}

 Variability in the $^{13}$CO and C$^{18}$O $J=2-1$ disk-integrated line-to-continuum flux ratios appears to arise outside of the $1\sigma$ uncertainties of the individual epochs (Figure~\ref{fig:obs}) but with no clear pattern.
 We find that the reduced $\chi^{2}$ compared to the mean ($\chi_{red}^{2}$) is $\sim$2 for the $^{13}$CO and C$^{18}$O line-to-continuum flux ratios, indicating intrinsic variability or that the uncertainties are slightly underestimated.
 We use Pearson's correlation to test if there is a linear relationship between the $^{13}$CO and C$^{18}$O line-to-continuum ratios and the Spearman rank correlation to measure the degree of correlation between the $^{13}$CO and C$^{18}$O line-to-continuum ratios.
 As shown in Figure~\ref{fig:corrALMA}, there is no strong correlation between the $^{13}$CO and C$^{18}$O line-to-continuum ratios. This is confirmed with the measured Pearson correlation coefficient of $\rho_p=0.3$ and its $p$-value of $p_p=0.2$. Likewise, the Spearman correlation coefficient ($\rho_s=0.4$) and its $p$-value ($p_s=0.1$) and the Kendall correlation coefficient ($\tau_k=0.3$) and its $p$-value ($p_k=0.1$) do not suggest correlated variability. A correlation would be expected if intrinsic variability affected both of the lines, whether by chemical (e.g., molecular dissociation) or physical (e.g., heating) means. This suggests that the observed changes are not due to intrinsic variability and instead that the uncertainties of the individual epochs are slightly underestimated, making the variations of the ALMA $^{13}$CO and C$^{18}$O line observations consistent with noise.
 
%FIGURE 
%%%%%%%%%%%%%%%%%%%%%%%%%%
%%%%%%%%%%%%%%%%%%%%%%%%%%

\subsubsection{3 cm Continuum Emission with VLA}

%FIGURE 
%%%%%%%%%%%%%%%%%%%%%%%%%%
\begin{figure}    
\epsscale{2.2}
\plottwo{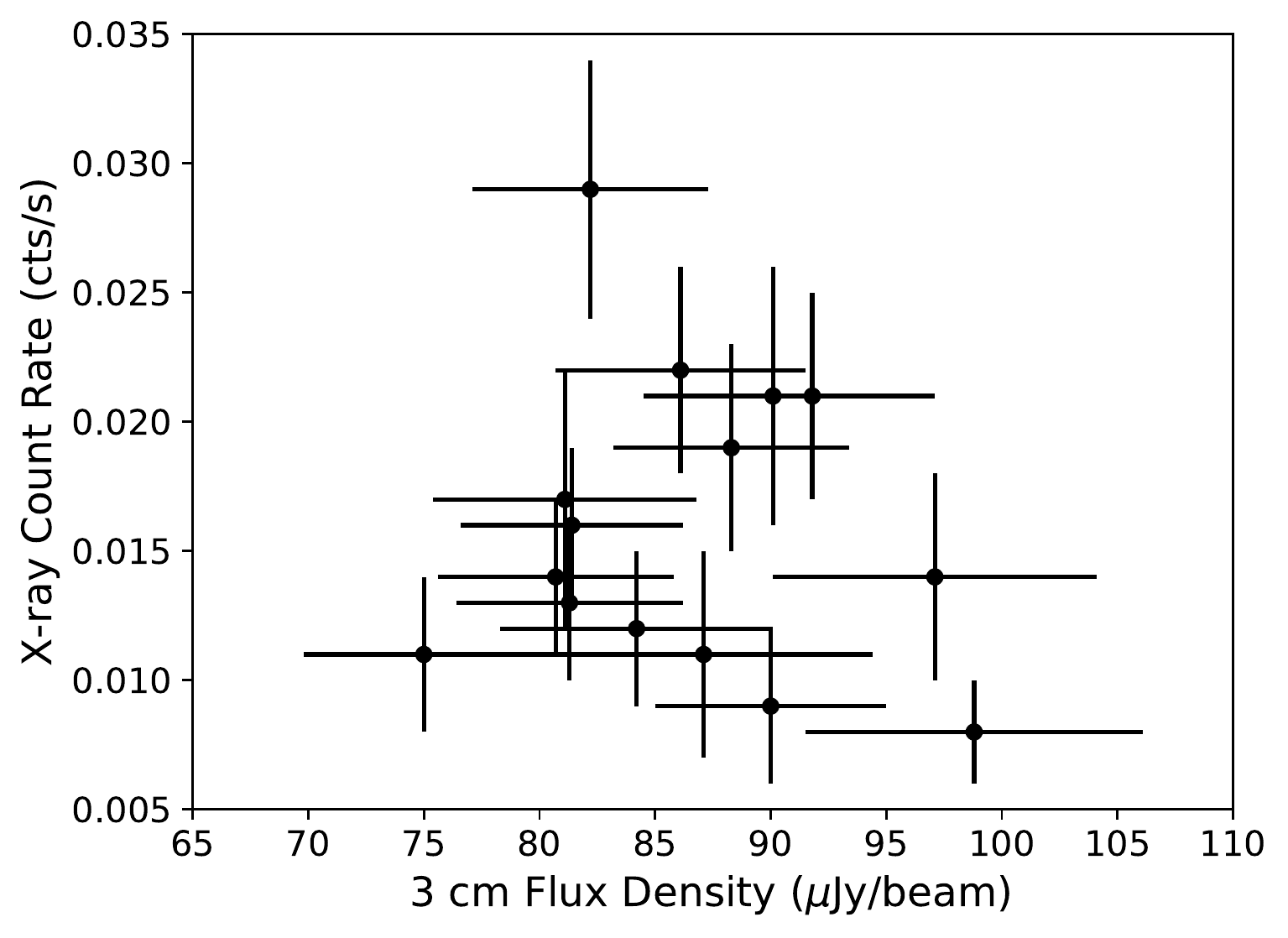}{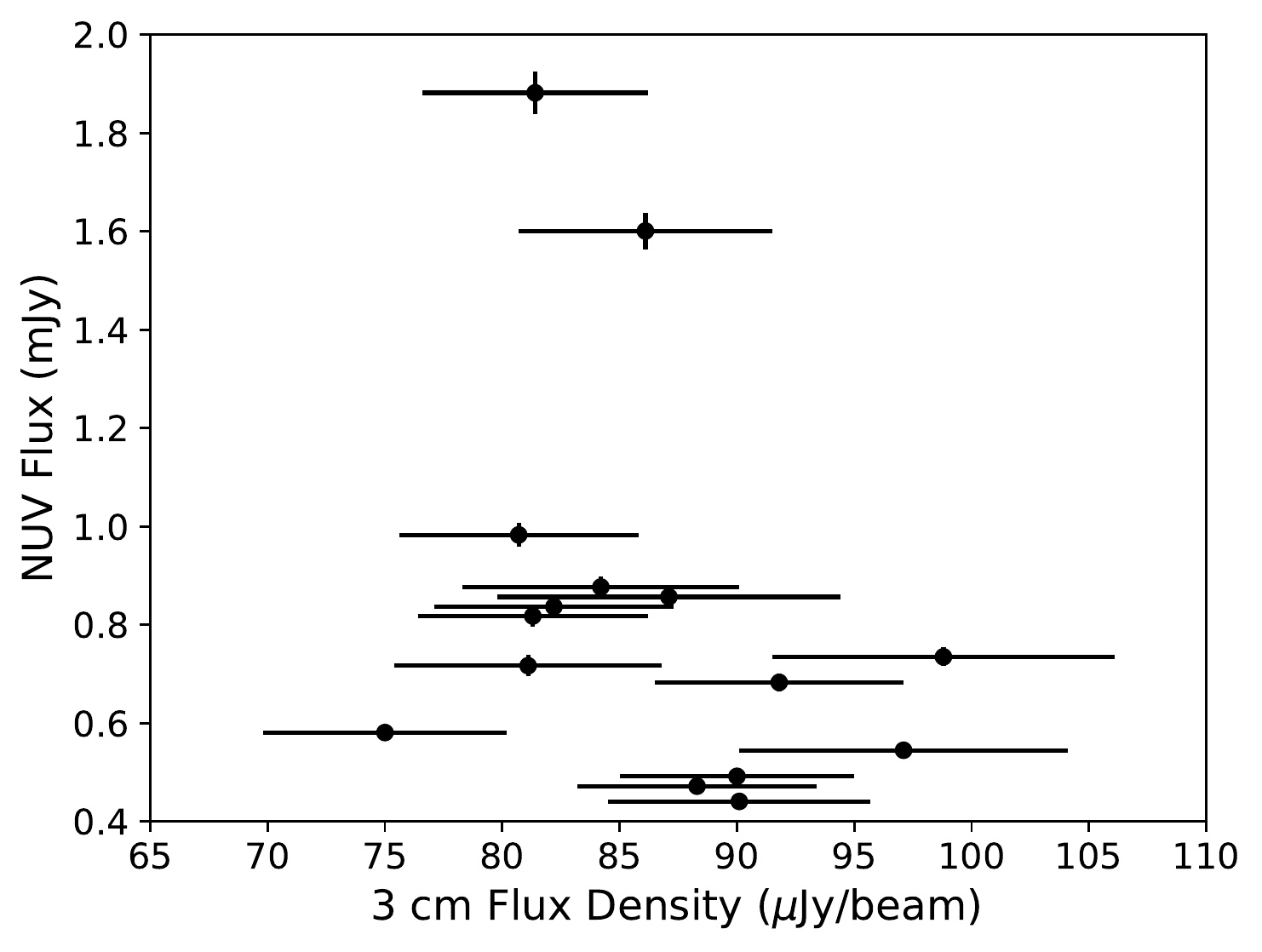}
\caption{
Comparisons of the {\it Swift} X-ray count rate and NUV flux from \citet{espaillat21}, with the 3~cm continuum flux density. We find no correlation between the 3~cm continuum emission and X-ray count rate ($\rho_p=-0.1$, $p_p=0.6$, $\rho_s=-0.03$, $p_s=0.9$, $\tau_k=-0.03$, $p_k=0.8$) or the 3~cm continuum emission and NUV fluxes ($\rho_p=-0.3$, $p_p=0.3$; $\rho_s=-0.4$, $p_s=0.1$; $\tau_k=-0.2$, $p_k=0.2$).
}
\label{fig:corrVLA}
\end{figure} 
%%%%%%%%%%%%%%%%%%%%%%%%%%

Figure~\ref{fig:obs} shows the 3~cm continuum flux densities and uncertainties measured at the VLA (Table~3). Also shown are the mean flux density and the mean uncertainties. We measure $\chi_{red}^{2}$$\sim$1, indicating that the observations and their uncertainties are consistent with a constant flux equal to the mean.  

There are no clear correlations with any of the other datasets here or in \citet{espaillat21}. In particular, we compare 15 values of the 3~cm continuum emission with the {\it Swift} X-ray and NUV fluxes from \citet{espaillat21} that were taken on the same days since these may be linked to photoionized emission and the mass-loss rate in the jet, respectively (Figure~\ref{fig:corrVLA}). We report the Pearson correlation coefficient ($\rho_p$) and its $p$-value ($p_p$), the Spearman correlation coefficient ($\rho_s$) and its $p$-value ($p_s$), and the Kendall correlation coefficient ($\tau_k$) and its $p$-value ($p_k$) in Figure~\ref{fig:corrVLA} and find no correlations. 

%FIGURE 
%%%%%%%%%%%%%%%%%%%%%%%%%%
\begin{figure*}    %figure 1
\epsscale{1.2}
\plotone{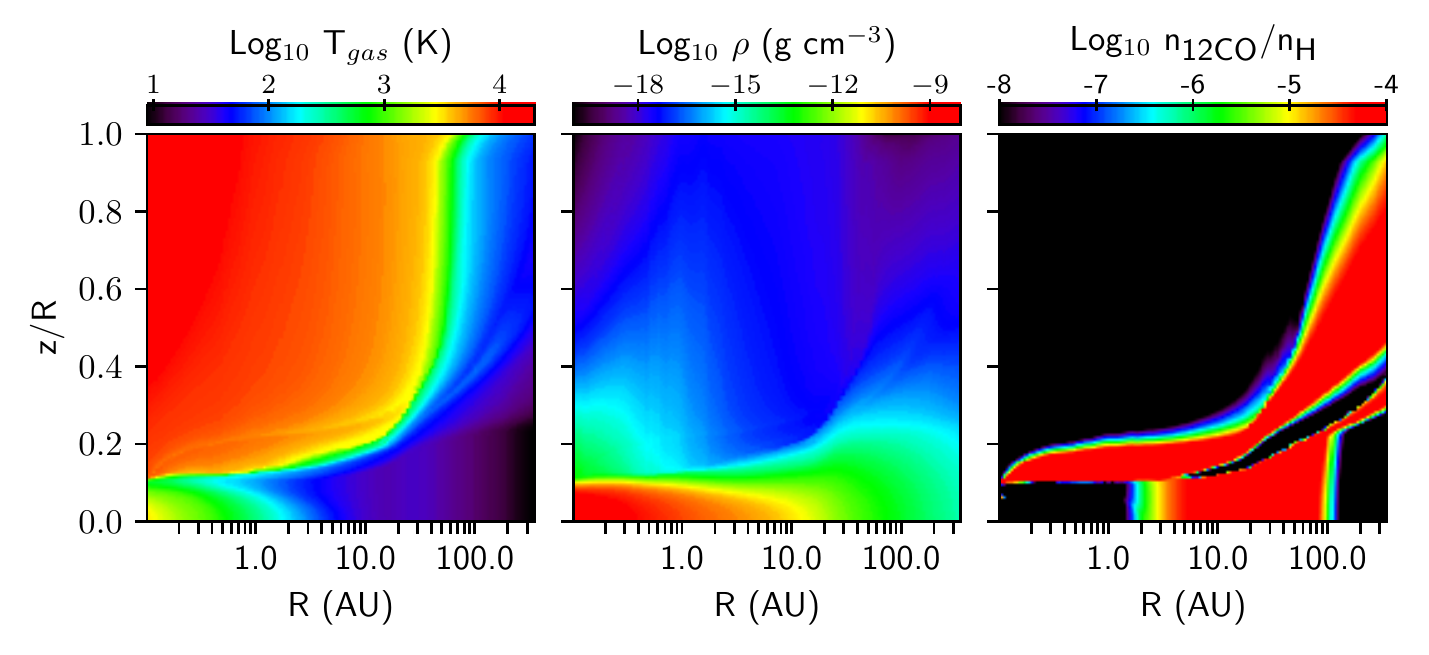}
\caption{
Base model of the gas structure of GM~Aur (i.e., before any changes in L$_{FUV}$ or L$_{X}$). We show the gas temperature (left), density (middle), and the CO$/$H abundance (right). Here we show the $^{12}$CO abundance; the $^{13}$CO and C$^{18}$O abundances are scaled by the abundances of $^{13}$C and $^{18}$O.  Note that the decrease in abundance at $\sim200$~au is due to CO freezing out onto grains (i.e., the snow surface). The very-low-density areas colored in black $<1$~au and the strip above the midplane from $\sim10$--100~au are due to carbon being ``locked'' in hydrocarbons, namely CH$_3$.
}
\label{fig:gasmodel1}
\end{figure*} 
%%%%%%%%%%%%%%%%%%%%%%%%%%

\subsection{Modeling Results}

In order to explore theoretical expectations for changes in the submillimeter CO emission lines, we create chemical models focusing on the $^{13}$CO and C$^{18}$O emission lines.   We study the effect of changes to the high-energy radiation fields: the FUV due to accretion and the X-ray due to accretion and/or the corona. 

\subsubsection{Modeling Framework} 

We used the D'Alessio irradiated accretion disk models \citep[DIAD,][]{dalessio06}, expanded to calculate the temperature of the gas and the dust separately. We assume that the disk is in hydrostatic equilibrium in the vertical direction and in centrifugal equilibrium in the radial direction. The disk is permeated by radiation from the star and the accretion shock. Dust settling and heating are treated as in \cite{dalessio06}. For the gas heating, we include viscous dissipation \citep{dalessio98}, photoelectric heating \citep{weingartner01}, photodissociation and ionization by X-rays and cosmic rays \citep{glassgold04,umebayashi09}, collisional de-excitation of H$_2$ \citep{kamp01}, and grain formation of H$_2$ \citep{kamp01}. Gas cooling is due to Ly $\alpha$, lines of [OI], [CII], [Ne II], CO, collisional excitation of H \citep{kamp01}, and gas-grain collisions \citep{glassgold04}. We implemented a reduced version of the time-dependent chemical model of \citet{fogel11}, with 273 reactions that include all the carbon chemistry and hydrocarbons up to CH$_3$, but no N or S networks, and some grain-gas interaction. We assume isotope ratios of $^{12}$C$/$$^{13}$C$\sim$89 and $^{18}$O$/$$^{16}$O$\sim$490.
The radiative transfer is carried out separating the direct and the diffuse radiation fields, following \citet{nomura07}. We include self-shielding of H$_2$ following \citet{draine96} and CO following \citet{visser09}. The density and temperature structures are obtained self-consistently by enforcing hydrostatic equilibrium.
Once we have the populations of the chemical species in the disk, we use the LIME code \citep{brinch10} to calculate the flux of the $^{13}$CO and C$^{18}$O lines. 

Our base model is based on the DIAD disk model that fits the SED and millimeter continuum images of GM~Aur in \citet{macias18}. This model uses an {\mdot} onto the star of 0.5$\times$10$^{-8}$~{\msunyr} and X-ray luminosity, L$_{X}$, of $4.4\times10^{30}$ erg~s$^{-1}$, consistent with the lower range of these values observed in GM~Aur by \citet{robinson19} and \citet{espaillat19a} and \citet{espaillat21}, respectively. We also use a maximum grain size (a$_{max}$) of 2~{\micron}, a dust settling parameter \citep[$\epsilon$; see][]{dalessio06} of 0.7, a disk radius of $\sim340$~au, and initial CO abundance relative to H of 10$^{-4}$. We note that we only use small grains in our model since here we are interested in the CO emission, which comes from the disk atmosphere. The base model starts with the initial abundances from \citet{cleeves14}.

We show the gas temperature (T$_{gas}$), density ($\rho$), and CO$/$H abundance of the base model in Figure~\ref{fig:gasmodel1}. This model gives a $^{13}$CO flux of 4.94~mJy \kms, consistent with the mean of the observations ($4.9\pm0.6$~mJy \kms) and a C$^{18}$O flux of 1.94~mJy \kms, roughly consistent with the mean of the observations ($1.1\pm0.3$~mJy \kms). We note that isotope-selective photodissociation is not included in our model, which may explain why the model has a slightly higher C$^{18}$O flux than the observed mean value.  The observed uncertainties listed here take into account both the measurement uncertainty noted in Table~\ref{Tab:ALMAflux} and the 5\% flux calibration uncertainty. 

\subsubsection{L$_{FUV}$ Increase Model} 

\begin{figure*}  
\epsscale{0.9}
\plotone{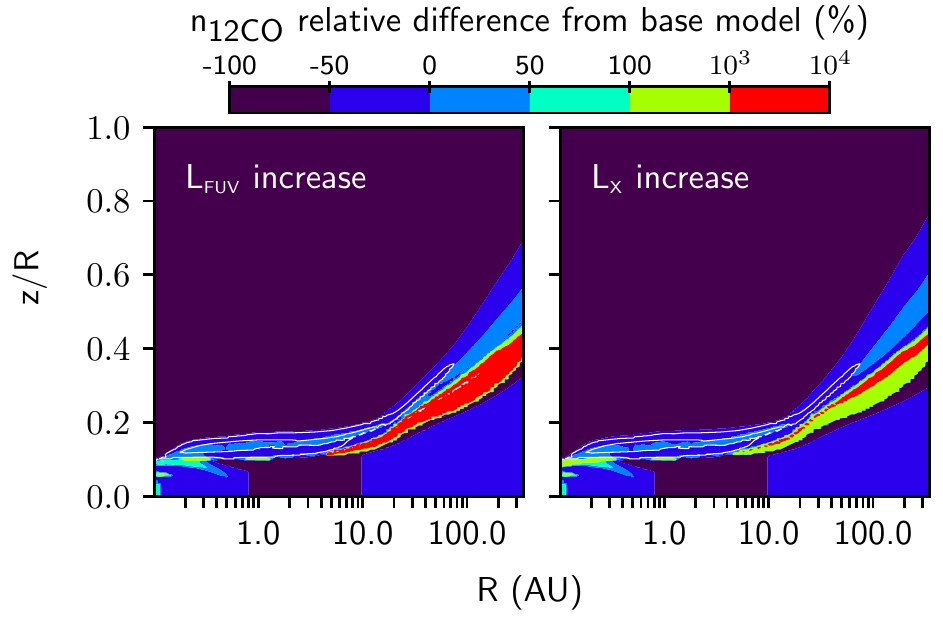}
\caption{
The change in CO/H abundances relative to the base model for an L$_{FUV}$ increase (left) and L$_{X}$  increase (right).  The base model is at time=0 and the length of both increases is 1~Myr. There are noticeable changes from the base model in Figure~\ref{fig:gasmodel1} which correspond to a $\sim6\%$ and $\sim25\%$ decrease in the $^{13}$CO line emission predicted by the L$_{FUV}$ and L$_{X}$ increase models (Table~\ref{tab:model}) and a $\sim14\%$ and $\sim36\%$ decrease in the C$^{18}$O line emission predicted by the L$_{FUV}$ and L$_{X}$ increase models (Table~\ref{tab:modelC18O}). 
The contour shows the region of higher density $^{12}$CO gas with $n_{\mbox{12CO}}>10^{-4}n_{\mbox{H}}$.
It is the negative relative difference in this region which causes the decrease in $^{13}$CO and C$^{18}$O line emission.}
\label{fig:gasmodel2}
\end{figure*}

To estimate the response of the $^{13}$CO and C$^{18}$O population to a change in FUV luminosity, L$_{FUV}$, the base model structure was irradiated with an enhanced L$_{FUV}$ for an extended period of time. The chemistry is run for the same duration of time that the disk is irradiated by the increased luminosity. The FUV increase has a luminosity consistent with $\dot {\rm M}=1.8\times10^{-8}$~\msunyr, the upper range of the accretion rates seen in GM~Aur to date \citep{robinson19}. Models were run with periods of increased luminosity of 1~d, 6~d, 1~yr, 100~yr, 1~kyr, 10~kyr, 100~kyr, 500~kyr, and 1~Myr. We note in Section~4.1 that the longer timescales are not realistic periods of luminosity enhancement for TTSs, but we include them here to explore the parameter space. We show the CO/H abundance after 1~Myr of a sustained L$_{FUV}$ increase and the relative change from the base model in Figures~\ref{fig:gasmodel2} (left panels) and list simulated $^{13}$CO and C$^{18}$O fluxes calculated with LIME in Tables~\ref{tab:model} and~\ref{tab:modelC18O}. This was computed in non-LTE with collisional rates taken from the LAMDA database \citep{schoier05}. The models indicate that the L$_{FUV}$ increase leads to a slight decrease in the $^{13}$CO and C$^{18}$O emission starting at 100~yr, which continues through 1~Myr to about 6$\%$ (Table~\ref{tab:model}) for $^{13}$CO and about 14$\%$ (Table~\ref{tab:modelC18O} for C$^{18}$O). This decrease is due to enhanced CO photodissociation by the FUV.  

%FIGURE 
%%%%%%%%%%%%%%%%%%%%%%%%%%
%%%%%%%%%%%%%%%%%%%%%%%%%%

\subsubsection{L$_{X}$ Increase Model}

\begin{figure}    
\epsscale{2.5}
\plottwo{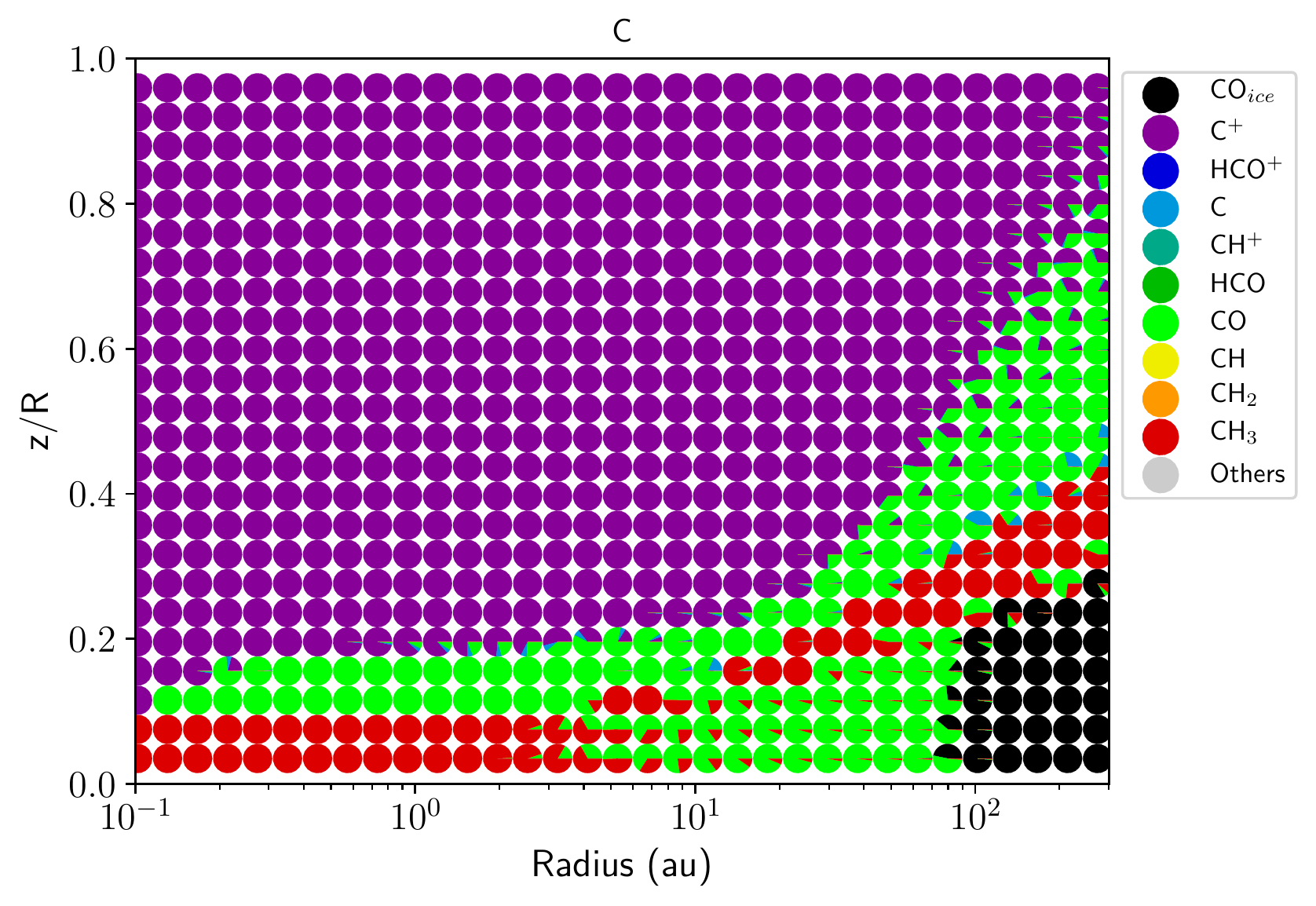}{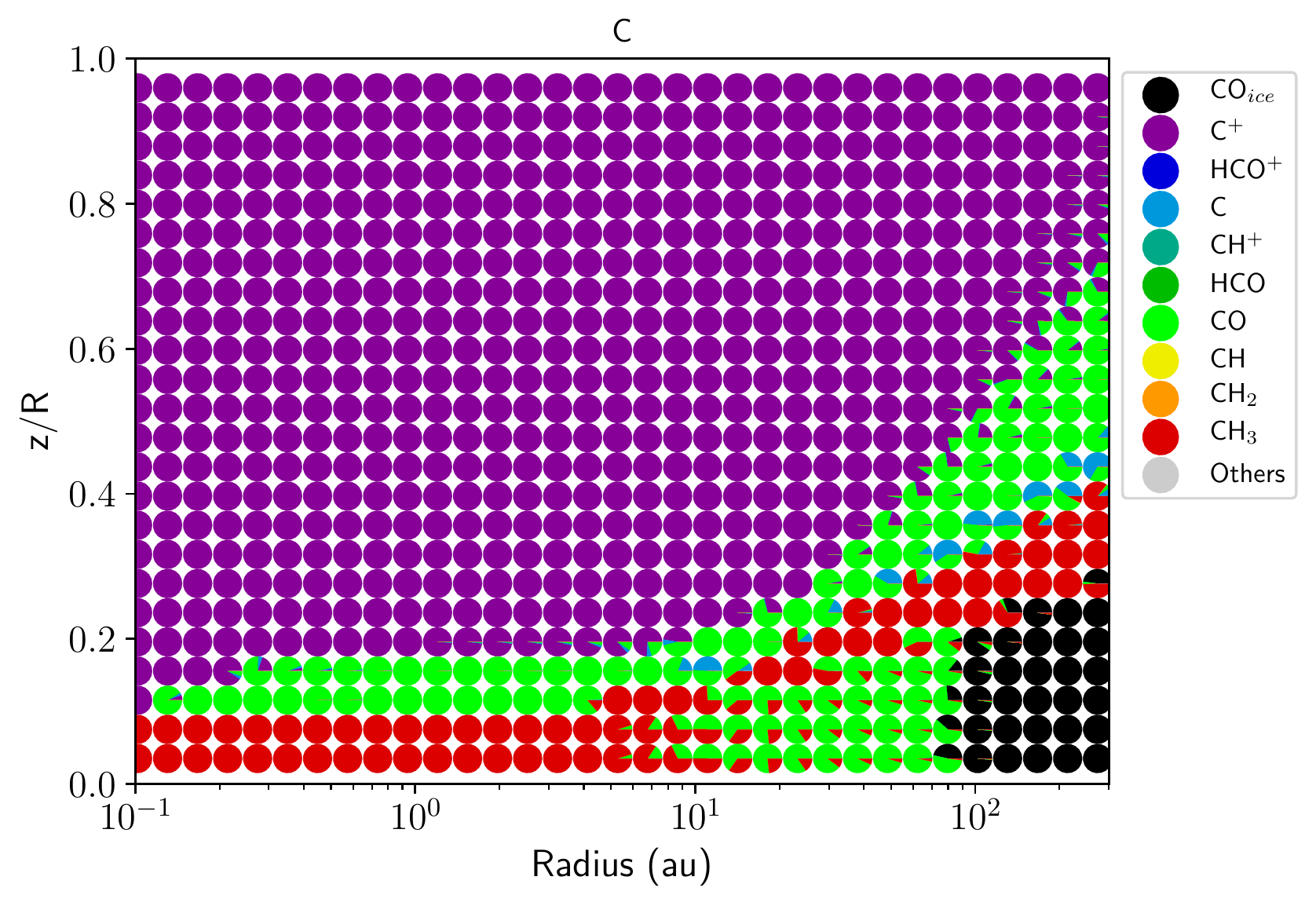}
\caption{
Comparison of the distribution of C for the base model at time=0 (top) and the L$_{X}$ increase model over 1~Myr (bottom). Each circle is a pie-chart of the main carbon carriers in each part of the disk which are listed in the legend. CH$_{3}$ should be interpreted as carbon being in hydrocarbons. In the L$_{X}$ increase model, some CO (green) in the upper layers is now in C+ (purple). There is also an increase in C (blue).
}
\label{fig:modelxray}
\end{figure}

To estimate the response of the $^{13}$CO and C$^{18}$O population to a change in L$_{X}$, the base model structure was irradiated with an increase of L$_{X}$ consistent with the upper range of L$_{X}$ ($1.7\times10^{31}$~erg~s$^{-1}$) observed in GM~Aur \citep{espaillat19a} and the same time lengths for the increased luminosity as noted above for the L$_{FUV}$ increase case. The L$_{X}$ does not change significantly over the period of our ALMA observations \citep{espaillat21}.  However, X-ray flares can last less than a day \citep{favata05} and the cadence of the {\it Swift} observations was daily, so a flare that was not observed at its peak luminosity could have occurred. Motivated by a change in the X-ray count rate about two days before the ALMA observations began \citep{espaillat21} and that \citet{espaillat19a} observed a factor of $\sim4$ change in the L$_{X}$ of GM~Aur, we test if changes in the X-ray could lead to observable variation in $^{13}$CO and C$^{18}$O line emission. We show the CO$/$H abundance for the L$_{X}$ increase case and the relative change from the burst model in Figures~\ref{fig:gasmodel2} (right panels) and list simulated $^{13}$CO and C$^{18}$O fluxes in Tables~\ref{tab:model} {and~\ref{tab:modelC18O})}. There is a $\sim25\%$ change in the $^{13}$CO line emission and a $\sim36\%$ change in the C$^{18}$O line emission when the L$_{X}$ is continuously higher for 1~Myr, higher than that seen from increasing the L$_{FUV}$, but the timescales to see changes are still long, at least 100~yr. In Figure~\ref{fig:modelxray}, we show the elemental C distribution in the disk in the base model at time=0 (top) and the L$_{X}$ increase model after 1~Myr (bottom).  
The L$_{X}$ increase model displays a decrease in CO abundance, with higher abundances of C+ and C. An increase in the L$_{X}$ can lead to an increase in HCO$^+$ formation via the formation of H$_3^+$ so long as the CO is not being destroyed more rapidly than H$_3^+$ is being enhanced \citep{cleeves17}.

%TABLE 
\begin{deluxetable}{c|cc|cc}  
\tablecaption{Simulated $^{13}$CO Line Fluxes  \label{tab:model}}
\tablehead{
\colhead{Increase} \vspace{-0.2cm} & \multicolumn2c{L$_{FUV}$ increase} & \multicolumn2c{L$_{X}$ increase} \\
\colhead{Length} \vspace{-0.2cm} & \colhead{Line Flux} & \colhead{\% change} & \colhead{Line Flux} & \colhead{\% change} \\
\colhead{} \vspace{-0.2cm} & \colhead{(mJy km~s$^{-1}$)} & \colhead{from base} & \colhead{(mJy km~s$^{-1}$)} & \colhead{from base} 
}
\startdata
%----------------------------------------------------------------------
0$^{a}$      & 4.94 & ...  &4.94 & ...\\
1 d             & 4.94 & 0    &4.97 &+0.6\\
6 d             & 4.93 & -0.2 &4.92 &-0.4\\
1 yr            & 4.92 & -0.4 &4.91 &-0.6\\
100 yr          & 4.74 & -4.0 &4.86 &-1.6\\
1 kyr           & 4.69 & -5.1 &4.82 &-2.4\\
10 kyr          & 4.70 & -4.8 &4.84 &-2.0\\
100 kyr         & 4.78 & -3.2 &4.63 &-6.3\\
500 kyr         & 4.77 & -3.4 &4.05 &-18.0\\
1 Myr           & 4.65 & -5.9 &3.73 &-24.5
%----------------------------------------------------------------------
\enddata
\tablenotetext{a}{This corresponds to the base model which is at time=0.}
\end{deluxetable} 

%TABLE 
\begin{deluxetable}{c|cc|cc}  
\tablecaption{Simulated C$^{18}$O Line Fluxes}  \label{tab:modelC18O}
\tablehead{
\colhead{Increase} \vspace{-0.2cm} & \multicolumn2c{L$_{FUV}$ increase} & \multicolumn2c{L$_{X}$ increase} \\
\colhead{Length} \vspace{-0.2cm} & \colhead{Line Flux} & \colhead{\% change} & \colhead{Line Flux} & \colhead{\% change} \\
\colhead{} \vspace{-0.2cm} & \colhead{(mJy km~s$^{-1}$)} & \colhead{from base} & \colhead{(mJy km~s$^{-1}$)} & \colhead{from base} 
}
\startdata
%----------------------------------------------------------------------
0$^{a}$      & 1.94 & ...  &1.94 & ...\\
1 d             & 1.94 & 0    &1.94 &0\\
6 d             & 1.93 & -0.5 &1.92 &-1.0\\
1 yr            & 1.94 & 0 &1.94 &0\\
100 yr          & 1.91 & -1.5 &1.95 &0.5\\
1 kyr           & 1.90 & -2.1 &1.91 &-1.5\\
10 kyr          & 1.89 & -2.6 &1.90 &-2.1\\
100 kyr         & 1.91 & -1.5 &1.78 &-8.24\\
500 kyr         & 1.76 & -9.3 &1.44 &-25.8\\
1 Myr           & 1.67 & -13.9 &1.24 &-36.1
%----------------------------------------------------------------------
\enddata
\tablenotetext{a}{This corresponds to the base model which is at time=0.}
\end{deluxetable} 

%FIGURE
%%%%%%%%%%%%%%%%%%%%%%%%%%
%%%%%%%%%%%%%%%%%%%%%%%%%%

\subsubsection{Observability} 

\begin{figure}    
\epsscale{1.2}
\plotone{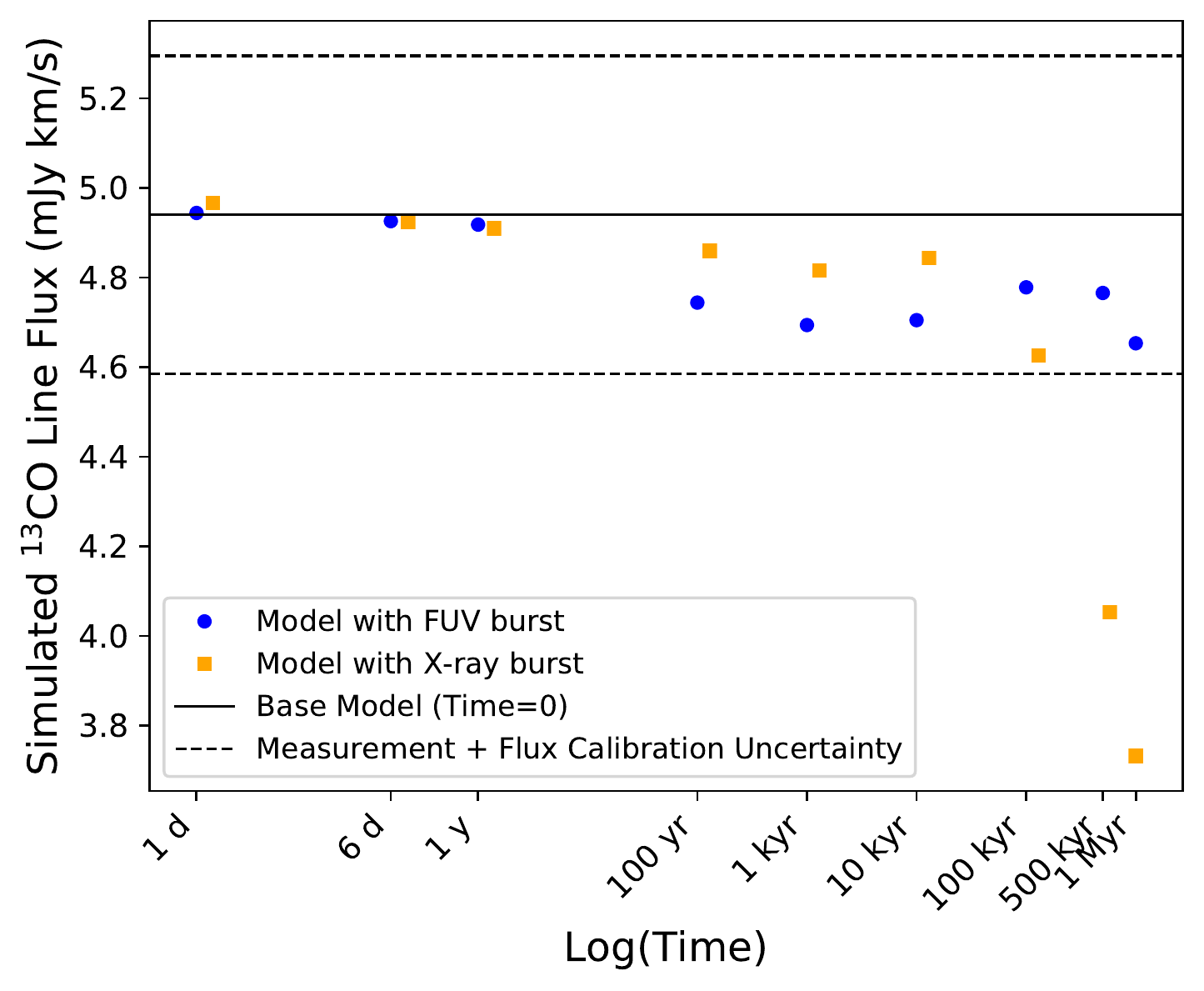}
\caption{
Simulated $^{13}$CO line fluxes. The solid line is the base model flux and the dotted lines correspond to the uncertainties from the ALMA measurement uncertainty and flux calibration uncertainty. We show $^{13}$CO line fluxes from the model (Table~\ref{tab:model}) with an L$_{X}$ increase (orange) and L$_{FUV}$ increase (blue). We see no observable change in the predicted line flux compared with the base model in the L$_{FUV}$ increase case and see a change in the L$_{X}$ increase case only after 500~kyr. (In order to avoid overlap with the L$_{FUV}$ increase points, the L$_{X}$ increase points have been slightly shifted to the right.)
}
\label{fig:gasmodel3}
\end{figure}

\begin{figure}    
\epsscale{1.2}
\plotone{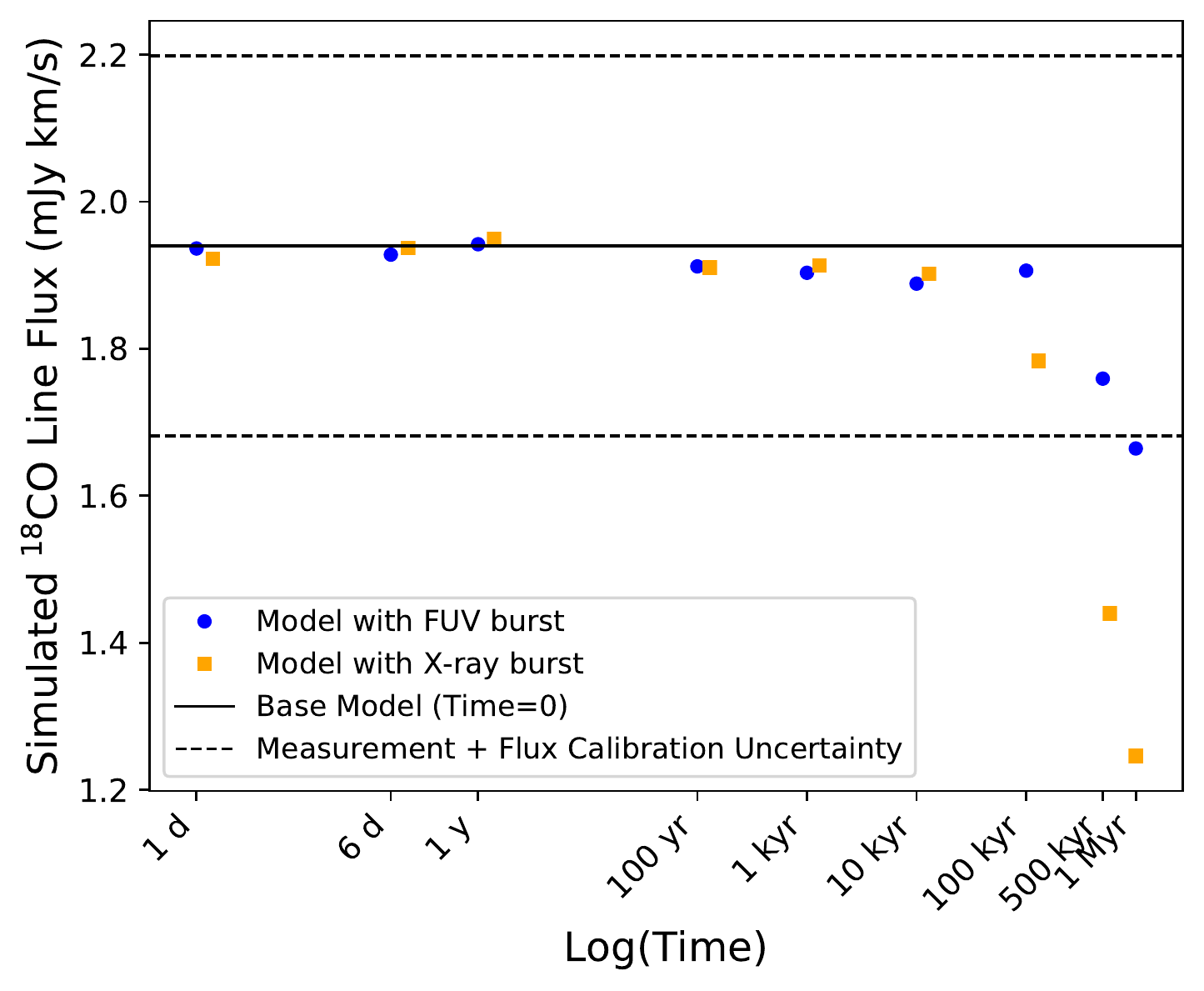}
\caption{
Simulated C$^{18}$O line fluxes. Lines and symbols are the same as Figure~\ref{fig:gasmodel3} and C$^{18}$O line fluxes are from Table~\ref{tab:modelC18O}. We see a change in the predicted line flux compared with the base model in the L$_{FUV}$ increase case after 1 Myr and see a change in the L$_{X}$ increase case after 500~kyr.
}
\label{fig:gasmodel3C18O}
\end{figure}

We explored whether the changes in the $^{13}$CO and C$^{18}$O line emission predicted by the disk chemistry model (Tables~\ref{tab:model} and~\ref{tab:modelC18O}) are detectable by ALMA. In Figures~\ref{fig:gasmodel3} and ~\ref{fig:gasmodel3C18O}, we plot the values in Tables~\ref{tab:model} and~\ref{tab:modelC18O} and compare them to the ALMA measurement and flux calibration uncertainties. We find that the changes predicted by the L$_{FUV}$ increase model (blue) are not observable for $^{13}$CO and only observable after 1~Myr for C$^{18}$O. For the L$_{X}$ increase model (orange), the changes predicted by the model would be observable only after 500~kyr for both $^{13}$CO and C$^{18}$O.  We note that ALMA's inherent flux uncertainties would hide the changes in CO line emission even if the integration time was increased. However, assuming that the line to continuum flux ratio can remove the effects of the absolute calibration errors, variability in CO emission could be detected. This supports our conclusion in Section~3.1.1 that the observed behavior of the $^{13}$CO and C$^{18}$O line emission is consistent with noise. We also discuss further in Section~4.1 that luminosity increases sustained on such long timescales are unlikely for TTSs.

%FIGURE 
%%%%%%%%%%%%%%%%%%%%%%%%%%
%%%%%%%%%%%%%%%%%%%%%%%%%%

%FIGURE 
%%%%%%%%%%%%%%%%%%%%%%%%%%
%%%%%%%%%%%%%%%%%%%%%%%%%%

\section{Discussion} \label{sec:discussion}

\subsection{The Impact of Variable FUV and X-ray Radiation on $^{13}$CO and C$^{18}$O Radio Emission}

We conclude that typical variability in L$_{FUV}$ and L$_{X}$ in CTTSs does not lead to observable variation in the commonly observed $^{13}$CO and C$^{18}$O emission lines based on the predicted magnitudes of the changes and the timescales necessary to see changes. DIAD models incorporating a self-consistent calculation of the gas temperature and chemistry show that the magnitude of the predicted changes in the $^{13}$CO and C$^{18}$O line emission are too small to be robustly detected with current instrumentation. Another issue is that our chemical modeling shows it would take L$_X$ or L$_{FUV}$ luminosity increases of at least 100 yr to lead to changes in the radio $^{13}$CO and C$^{18}$O emission lines. Even the rapidly accreting FUors with their large outbursts are expected to sustain luminosity increases only for a couple of hundred years \citep{audard14}, making the length of the luminosity increase necessary to see radio CO emission line changes in CTTSs implausible. We also do not expect changes in the CO emission lines due to changes in disk temperature. The approximate heating timescales of our models are 25~years close to the midplane and about 10~years in the disk atmosphere, where most of the CO emission originates.

In addition to our modeling results discussed above, we can also discount that the small changes seen in the CO line-to-continuum ratios in the ALMA data are intrinsic to GM Aur since \citet{espaillat21} conclude that the accretion variability seen in GM~Aur is due to the rotational modulation of accretion (i.e., due to our viewing angle) and not to global changes in the accretion rate. Therefore, we would not expect to see changes in the gas emission lines traced by ALMA due to changes in L$_{FUV}$ in this campaign. While the observer sees a variable accretion rate and L$_{FUV}$ as the hot spot rotates into and out of view, some part of the disk is always being irradiated by the same accretion luminosity, and the ALMA observations trace emission from most of the disk.  

While we show that variability in $^{13}$CO and C$^{18}$O  emission lines for a CTTS would be very difficult to detect, disk chemistry changes due to variability in L$_{FUV}$ and L$_{X}$ may be detected at other wavelengths in CTTSs and for much larger luminosity increases than observed in CTTSs. Detecting variability on timescales of days due to X-ray variability appears promising by observing lines that are more sensitive to changes in the X-ray emission, such as HCO$^+$ in the radio \citep{cleeves17} and H$_2$O in the infrared \citep{waggoner19}. However, even in these cases, strong X-ray flares or long-duration weak flares are necessary to lead to observable changes. To further understand the impact of FUV and X-ray variability on disk chemistry in individual objects, EXors present another avenue of study. EXors undergo larger accretion bursts than CTTSs and on shorter timescales than FUors, making their luminosity increases more likely to be observed. Chemical modeling of these objects has shown that the large accretion bursts of EXors can evaporate CO ice in the envelope, leading to a detectable excess of gas-phase CO in the radio \citep{visser12,vorobyov13}.  Changes in H$_2$O and OH infrared line emission have been observed during an outburst in EX~Lup \citep{banzatti12}.

We conclude that the stellar and accretion variability typically seen in CTTSs does not have an important effect on the most commonly observed tracers of the outer disk. Changes in L$_{FUV}$ and L$_{X}$ may have an observable effect on tracers such as HCO$^+$ \citep[e.g.,][]{cleeves17}, but the unpredictable nature of variability makes it necessary to coordinate multiwavelength observations to ascertain a robust connection.

\subsection{The Accretion-Jet Connection}

Our data were taken in a low-resolution configuration of the VLA, and so the 3~cm continuum emission seen here in GM~Aur is due to both the jet and photoionized disk \citep{macias16}. Deciphering if increases in the 3~cm emission are due to the photoionized disk emission or the jet emission is in principle possible by studying the time delay between increases in the accretion rate and X-ray emission and increases in the 3~cm flux. A short delay (a few hours; e.g., 50~au is 7~light-hours) is expected between an increase in X-ray emission and gas photoionization. A longer delay (a few days) is expected for a correlation between accretion onto the star and ejection of material. Assuming a jet velocity of $\sim300$~\kms\ and that the gas in the jet is ionized by the shocks in the recollimation point at $\sim1$~au from the star, the time delay would be $\sim6$~days. If the recollimation point is at larger distances, the delay would be even longer. Here we do not see correlations with the X-ray or NUV emission (Figure~4) that would suggest a connection to changes in the photoionized disk emission or jet emission, respectively.

As a comparison, in 2018, \citet{espaillat19b} examined three epochs of simultaneous L$_{X}$, accretion rate, and 3~cm continuum measurements of GM~Aur and reported a change in the 3~cm continuum emission between the first and second epoch and linked this to accretion. These changes were attributed to the jet emission since the L$_{X}$ did not change and so the photoionized disk emission was not expected to have changed. \citet{espaillat19b} found that the magnitude of the observed change in the 3~cm continuum emission between the first and second epoch was consistent with the expected change from the observed correlation between the radio emission and outflow momentum rate \citep{anglada15}, given the observed accretion rates of GM~Aur. \citet{espaillat19b} noted that the changes in the mass-loss rate and mass accretion rate were not expected to be instantaneous but could be tracing an increase in mass accretion rate from several days earlier that was not observed.

While the three values for the 3~cm continuum flux in 2018 are outside of one another's uncertainties ($80.6\pm3.7~\mu$Jy, $95.2\pm5.4~\mu$Jy, $92.7\pm4.4~\mu$Jy), they are within the range of fluxes seen in 2019 (Figure~2, Table~3). One interpretation is that the magnitude of the observed changes being consistent with the correlation between the radio emission and outflow momentum rate (and hence accretion rate) was a coincidence. On the other hand, the {\mdot} measured in the second epoch of data in 2018 was the highest {\mdot} ever measured in 15 epochs of {\it HST} data of GM~Aur \citep[][]{bergin04,ingleby15,robinson19,espaillat21}. This 2018 epoch is also unique in that accretion columns with even higher densities were needed to fit the data, compared with any GM~Aur {\it HST} data taken to date \citep[][]{robinson19,espaillat21}. With the 2019 data, \citet{espaillat21} conclude that no global changes occurred in {\mdot} and that the changes seen were due to rotational modulation. It is possible that in addition to the rotational modulation, a strong accretion burst occurred in 2018, and the changes in the jet emission may be linked to this burst. It is difficult to confirm this since the expected changes in the 3~cm continuum are small and at the edge of what can be robustly detected given observational uncertainties. 

Detecting clear radio variability in CTTSs appears to be challenging, but other Young Stellar Objects (YSOs) are known to be variable at radio wavelengths. \citet{wendeborn20} found evidence for millimeter variability in an FUor object which could be explained by variability in the free-free emission due to changes in the jet/wind. Centimeter observations of embedded YSOs have revealed radio flux increases of a factor of 1.5 that appear to be correlated to accretion rates \citep{liu14}.  On the other hand, \citet{galvan-madrid15} found that a Class 0 YSO with a large accretion burst showed no evidence of a corresponding centimeter burst from the radio jet. FUors, EXors, and more embedded objects, all of which display much larger changes in accretion rate than CTTSs, may make better targets than CTTSs to explore the connection between mass accretion and mass ejection.

\section{Summary \& Conclusions} 

We present radio data from a multiwavelength variability campaign of the CTTS GM~Aur that spanned X-ray to centimeter wavelengths over a month in 2019.  We find no significant variability in the ALMA $^{13}$CO and C$^{18}$O line emission or the VLA 3~cm continuum emission. We show with chemical modeling that typical changes in the L$_{FUV}$ and L$_{X}$ of CTTSs do not lead to detectable changes in the millimeter $^{13}$CO and C$^{18}$O line emission. Accretion bursts may lead to changes in the mass-loss rate, and hence jet emission traced by the VLA 3~cm continuum emission, but this hypothesis would be difficult to confirm given the variable nature of accretion and measurement uncertainties. We conclude that the typical ranges of variability seen in the X-ray and FUV emission of CTTSs do not lead to observable changes in millimeter CO emission or centimeter continuum emission. Objects with more extreme variability, such as EXors, should be considered to further explore the connection between changes in high-energy radiation and radio emission.

\software{DIAD \citep{dalessio06}, CASA (v. 5.6.1-8; \citet{mcmullin07}, keplerian$\_$mask.py (Teague 2020), LIME \citep{brinch10}}\\

 \acknowledgments{We  thank  the  referee  for  a  constructive report.
This paper utilizes the D’Alessio irradiated accretion disk (DIAD) code. We wish to recognize the work of Paola D’Alessio, who passed away in 2013. Her legacy and pioneering work live on through her substantial contributions to the field. C.C.E. acknowledges support from HST grant GO-16010, NASA grant 80NSSC19K1712, and NSF Career grant AST-1455042.
N.C. acknowleges support from NASA grant NNX17AE57G.
This paper makes use of ALMA data ADS/JAO.ALMA$\#$2019.1.01437.S. ALMA is a partnership of ESO (representing its member states), NSF (USA) and NINS (Japan), together with NRC (Canada), MOST and ASIAA (Taiwan), and KASI (Republic of Korea), in cooperation with the Republic of Chile. The Joint ALMA Observatory is operated by ESO, AUI/NRAO, and NAOJ.
The National Radio Astronomy Observatory is a facility of the National Science Foundation operated under cooperative agreement by Associated Universities, Inc.
}

\appendix

\begin{figure}    
\epsscale{1.0}
\plotone{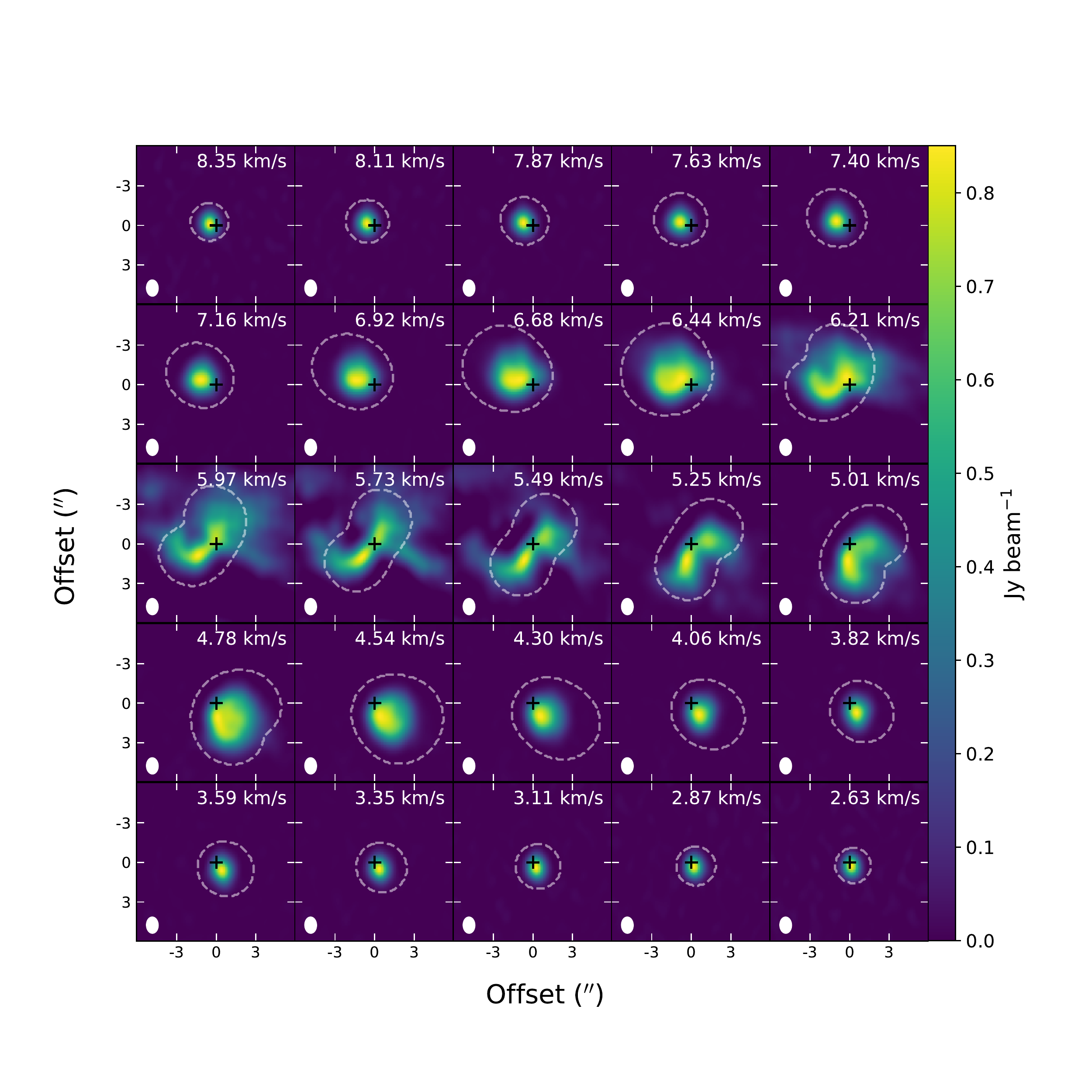}
\caption{
Channel maps for the $^{12}$CO line emission. The central black `+' denotes the location of GM Aur and the dashed lines mark the Keplerian mask. The average beam is shown in the lower left of each panel.
}
\label{fig:appendix1}
\end{figure}

\begin{figure}    
\epsscale{1.0}
\plotone{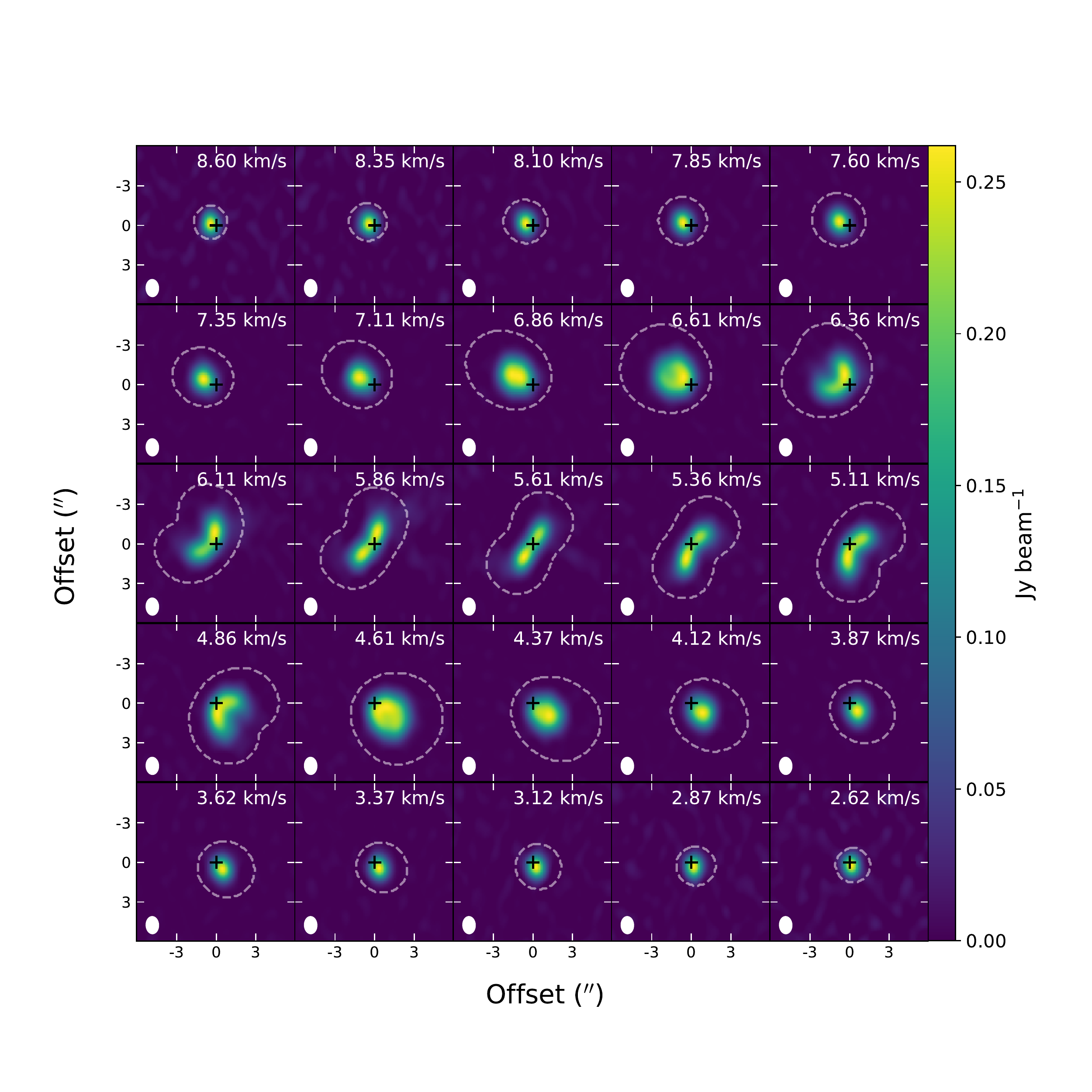}
\caption{
Channel maps for the $^{13}$CO line. Symbols are the same as in Figure~\ref{fig:appendix1}. 
}
\label{fig:appendix2}
\end{figure}

\begin{figure}    
\epsscale{1.0}
\plotone{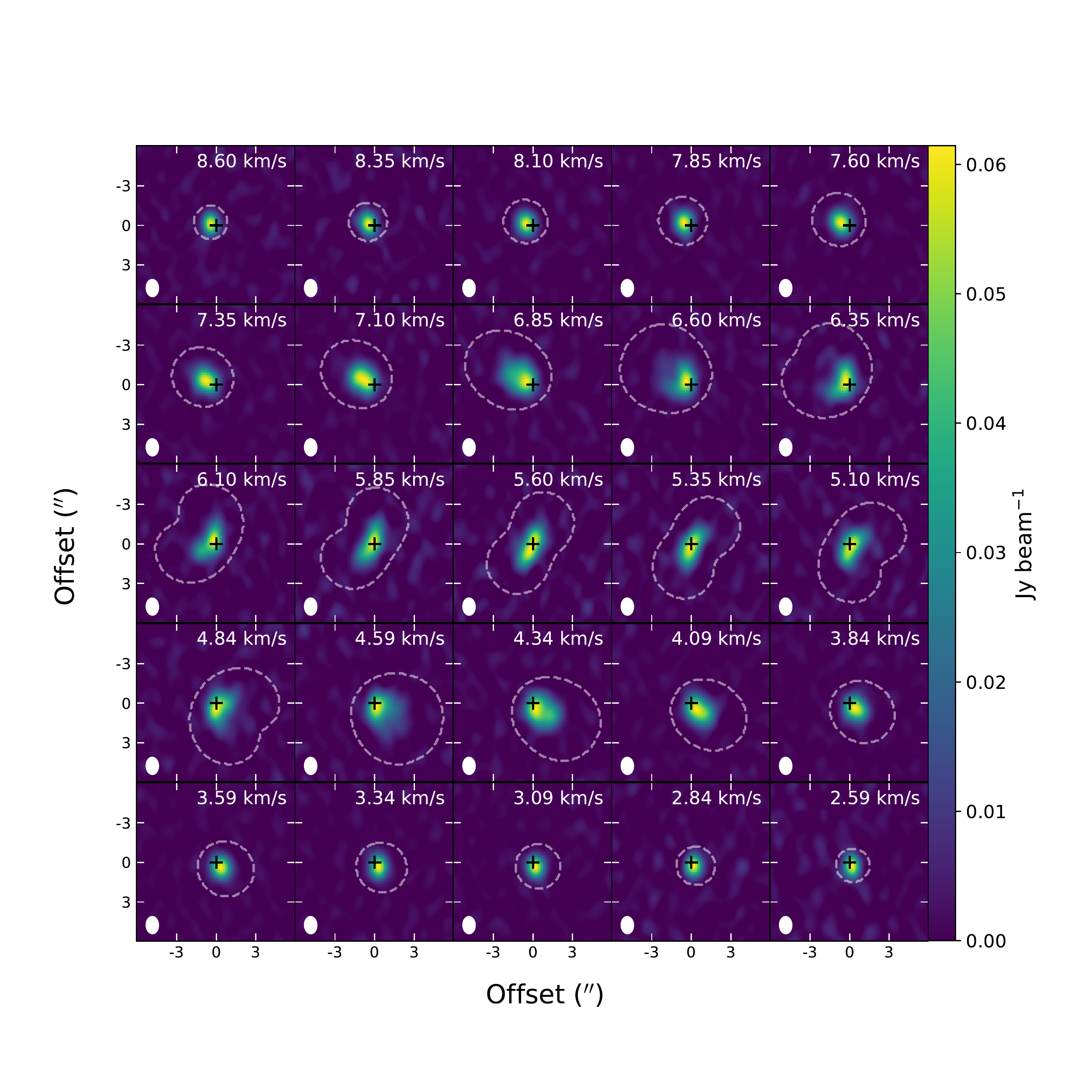}
\caption{
Channel maps for the C$^{18}$O line emission. Symbols are the same as in Figure~\ref{fig:appendix1}. 
}
\label{fig:appendix3}
\end{figure}

We show the channel maps of the $^{12}$CO, $^{13}$CO, and C$^{18}$O emission in Figures~\ref{fig:appendix1}, \ref{fig:appendix2}, and~\ref{fig:appendix3}, respectively.  Here we combine all the epochs.  The data were first averaged by three channels to emphasize the emission, and the resulting channel widths for $^{12}$CO, $^{13}$CO, and C$^{18}$O are 0.120, 0.125, and 0.125 km/s, respectively. Only emission within the Keplerian mask contributed to the moment-0 maps and flux estimates. 

%FIGURE 
%%%%%%%%%%%%%%%%%%%%%%%%%%
%%%%%%%%%%%%%%%%%%%%%%%%%%

%FIGURE 
%%%%%%%%%%%%%%%%%%%%%%%%%%
%%%%%%%%%%%%%%%%%%%%%%%%%%

%FIGURE 
%%%%%%%%%%%%%%%%%%%%%%%%%%
%%%%%%%%%%%%%%%%%%%%%%%%%%

\bibliographystyle{aasjournal}

\end{document}